\documentclass[aps,a4paper,pre,eqsecnum,showpacs,floatfix,twocolumn,merge,longbibliography,elide]{revtex4-1}

\usepackage{amsmath}
\usepackage{amsfonts}		
\usepackage[usenames,dvipsnames,svgnames]{pstricks}		

\usepackage{graphicx}		
\usepackage{tikz}			
\usepackage{tkz-graph}		

\usepackage[charter,expert]{mathdesign}				

\usepackage[ulem=normalem]{changes}
\setaddedmarkup{\blue{#1}}
\setdeletedmarkup{\red\sout{#1}}

\newcommand{\REM}[1]{}

\usepackage[%
	  unicode=true,%
	  colorlinks=true,%
	  linkcolor=blue,anchorcolor=blue,%
	  breaklinks=true,%
	  citecolor=blue,filecolor=cyan,%
	  menucolor=blue,%
	  breaklinks=true,%
	  urlcolor=blue]{hyperref}

\tikzset{
  EdgeStyle/.append style = {<->, bend left=45, line width = 0.05cm, 
  shorten <=1pt, shorten >=1pt} 
	  }	
	
\begin{document}
	
\title{Magnetic Exchange in Spin Clusters}
\author{M.~Georgiev}
\email{mgeorgiev@issp.bas.bg}
\affiliation{Institute of Solid State Physics, Bulgarian Academy of Sciences,
Tsarigradsko Chauss\'ee 72, 1784 Sofia, Bulgaria}
\author{H.~Chamati}
\affiliation{Institute of Solid State Physics, Bulgarian Academy of Sciences,
Tsarigradsko Chauss\'ee 72, 1784 Sofia, Bulgaria}

\date{\today}

\begin{abstract} 
We investigate the role of exchange bridges in molecular magnets.
We explore their effects on the
distribution of the valence electrons and their contribution to the
exchange processes. The present study is focused on a spin-half dimer
with nonequivalent exchange bridges. Here, we derive an effective
Hamiltonian that allows for an accurate estimation of the observables
associated to the magnetic
properties of the magnet. Our results are compared to those obtained
by means of the conventional
Heisenberg model that usually fails.
\end{abstract}
\maketitle

\section{INTRODUCTION}

Since decades the contribution of different bridging complexes 
to the magnetic 
properties of molecular magnets has motivated researchers to develop
different approaches and give suggestions
about the influence of
bridging ions between neighboring magnetic centers.
As the low-spin and short-bridged magnetic compounds are ideal candidates
to study magneto-structural features,
through the years, in many dimeric \cite{hay_1975,felthouse_1977,lorosch_1987} 
and trimeric \cite{gehring_1993} compounds the nature of bridging structure 
is the subject of constant debate.
One simple example is the Cu$^{2+}$ cubane-type complex
\cite{astheimer_1986} with symmetric bridges described in
the framework of a bilinear spin Hamiltonian.  
Further, the correlation of exchange constants with the structure parameters in 
alkoxo bridged cooper dimers was pointed out first
in Ref. \cite{astheimer_1986} and more recently in Ref. \cite{fraser_2017}.
Additional efforts relating the bridging complexes and magnetism
was shown
in Ref. \cite{charlot_1986}, where the interest in azido bridged complexes
continues for decades 
\cite{aebersold_1998,ribas_1993,ribas_1999,sadhu_2017,zhao_2017}. 
Other prominent examples of magneto-structural effects are the complexes
with Fe magnetic centers 
\cite{angaridis_2005,mekuimemba_2018,gregoli_2009,viennois_2010}, the Ni based 
compounds \cite{loose_2008, panja_2017,das_2017,woods_2017} and 
the Mn spin clusters \cite{goodenough_1955,defotis_1990,law_2000,han_2004,perks_2012,gupta_2016,hanninen_2018}.

In this paper we study the contribution of
bridging structures to the exchange 
processes and its effect on the magnetic spectrum.
In particular we show that complex bridging favors multiple
exchange pathways between magnetic centers.
To this end we discuss a spin-half dimer
consisting of two non symmetric exchange bridges.
We estimate the effect of both bridges and their contributions
by deriving an effective Hamiltonian that allows for discrete
exchange parameters accounting for the effects of bridging
ions. 
The proposed Hamiltonian can be applied
to a variety of molecular 
magnets with complex chemical environment and distortion 
in structure.
Recently \cite{georgiev_2018} we tested the proposed Hamiltonian 
by studying the 
magnetic spectra of trimeric compounds 
A$_3$Cu$_3$(PO$_4$)$_4$, A=(Ca, Sr, Pb) and the
Nickel tetramer spin cluster Ni$_4$Mo$_{12}$.

\section{THE GENERIC HAMILTONIAN}\label{sec:coulomb}
The generic Hamiltonian related to the electron-electron
and electron-nuclei interactions in a magnetic cluster, within the
adiabatic approximation, reads
\begin{equation}\label{eq:CoulombHamilton}
	\hat{H}=\sum_{i}\frac{\hat{p}^2_{i}}{2\mathrm{m}_i}+
	\sum_{\eta,i}\hat{U}(r_{\eta i})+\tfrac12\sum_{i\ne j}\hat{R}(r_{ij}),
\end{equation}
where $\hat{p}_{i}$ and $\mathrm{m}_i$ denotes the $i$-th electron
momentum operator and mass, respectively.
The potential energy operator 
$\hat{U}(r_{\eta i})$ accounts for the interaction of
the $i$-th electron with the $\eta$-th nucleus, separated by the
distance $r_{\eta i} = |\mathbf{r}_{i}-\mathbf{R}_{\eta}|$ with
$\mathbf{r}_i=(x_i,y_i,z_i)$ the coordinates of the $i$-th electron
and $\mathbf{R}_{\eta}$ those of the $\eta$-th nuclei.
The operator $\hat{R}(r_{ij})$ is related to the electrostatic repulsion 
between $i$-th and $j$-th electrons over the distance 
$r_{ij}=|\mathbf{r}_i-\mathbf{r}_j|$.
Obtaining the eigenstates of Eq. \eqref{eq:CoulombHamilton} 
assuming Coulomb potentials is a difficult task.
Therefore, we estimate the transition energy associated to the 
exchange processes with the aid of the variational technique described
in the following section.

\subsection{Molecular orbital approach}\label{ssec:molecular}
For complex bondings we consider the 
approach of delocalized electrons developed within the framework of molecular 
orbital (MO) theory \cite{fleming_molecular_2009}. 
Within MO theory the electrons are not  localized around the
constituent nuclei, but are rather 
distributed over the entire molecule, thus occupying
molecular orbitals. 
These orbitals are approximately given as a linear combination of the initial 
atomic orbitals (LCAOs) \cite{lennard_1929,mulliken_1939}.
Different constructions of molecular orbitals directly applied to study
exchange processes in dimer complexes can be found in Ref. \cite{hay_1975}. 
According to MO approach one distinguishes three categories of
orbitals according to their contributions to the bonding energy and
hence to the distribution of electrons. 
For further information on this topic we refer the interested
reader to Refs.
\cite{lennard-jones_1949,pople_1965,pople_1966,pople_1967,england_1971,reed_1985}
and references therein.

In constructing the molecular orbitals we assume that only 
the valence orbitals of the nearest coupled ions overlap.
Thus, we represent the $n$-th molecular orbital by the linear
combination
\begin{equation}\label{eq:MOrbitals}
	\phi_{n,m_i}(\mathbf{r}_i)=
	\sum_\eta c^\eta_n
	\psi^{\eta}_{\mu_\eta,m_i}(\mathbf{r}_i),
\end{equation}
where the real coefficients $c^\eta_n$ are functions of 
the overlap integrals between the directly coupled ions and
the electronic eigenstates in a potential created by the
$\eta$-th ion are given by
\begin{equation}\label{eq:AtomEigenstates}
	\psi^{\eta}_{\mu_\eta,m_i}(\mathbf{r}_i)
	=
	\prod_{\alpha\in\mathbb{K}}
	\psi^{\eta}_{\mu_\eta}(\alpha_i)\lvert m_i\rangle, \quad
	\mathbb{K}=\{x,y,z\},
\end{equation}
with $m_i$ the magnetic quantum number of the $i$-th 
electron. It is worth mentioning that
for all $\eta$ and $i$ the functions in Eq. \eqref{eq:AtomEigenstates} 
are orthogonal an normalized such that
\begin{equation}\label{eq:NormAtomState}
	\int
	\bar{\psi}^{\eta}_{\mu^{\phantom{a}}_\eta,m^{}_i}(\mathbf{r}_i)
	\psi^{\eta}_{\mu'_\eta,m'_i}(\mathbf{r}_i)\mathrm{d}\mathbf{r}_i
	= \delta_{\mu_\eta\mu'_\eta}\delta_{m^{}_i m'_i},	
\end{equation}
and the overlap integral is
\begin{equation*}\label{eq:AtomOverlap}
	0\le
	\int
	\bar{\psi}^{\eta^{}}_{\mu^{\phantom{a}}_{\eta},m_i}(\mathbf{r}_i)
	\psi^{\eta'}_{\mu'_{\eta'},m_i}(\mathbf{r}_i)
	\mathrm{d}\mathbf{r}_i
	<1.
\end{equation*}

Let us finally point out that the functions in Eq. \eqref{eq:MOrbitals} satisfy the orthogonality condition
\begin{equation}\label{eq:OrthogonalOrbitals}
	\int
	\bar{\phi}_{n,m^{}_i}(\mathbf{r}_i)
	\phi_{n',m'_i}(\mathbf{r}_i) 
	\mathrm{d}\mathbf{r}_i = 
	\delta_{nn'}
	\delta_{m^{}_im'_i}.
\end{equation}

\subsection{Dimer state functions}
Consider $N$ valence electrons, where $N\ge2$ is even. 
Let $n_k=N/2+k$ be the number of highest in energy occupied molecular orbital, 
where $k\in\mathbb{N}_{0}$.
As the transfer of electrons includes orbitals that are not
fully occupied,
in the spin-half dimer magnets the corresponding process requires two orbitals.
Therefore, knowing that $k$ is not fixed we assume these orbitals to
be $N/2+k$ and
$N/2+j$, where $k\ne j$. The remaining $(N-2)/2$ molecular orbitals are fully occupied.
As a consequence, we distinguish two sets of state functions.
These states are obtained taking into account the Slater determinant \cite{slater_1929,slater_1930} 
and symmetrizing the corresponding functions according to the spin quantum 
numbers of each electron pair. 
The first set of states describe fully occupied molecular orbitals.
These are
\begin{widetext}
\begin{subequations}\label{eq:DimerStateFunctions}
	\begin{align}\label{eq:OneOrbitalSymmetrized}
		\Phi^{n_k,n_j}_{S,M}(\mathbf{r}_1,\ldots,\mathbf{r}_{N}) =
		\sum_{P_{\mathbf{r}_1\ldots\mathbf{r}_{N}}}
		c_{\mathbf{r}_{N-1},\mathbf{r}_{N}}
		\prod_{i}^{\frac{N}{2}-1}
		\frac{
			\Phi^i_{s_{2i-1,2i}}(\mathbf{r}_{2i-1},\mathbf{r}_{2i}) 
		}{\sqrt{2^{\frac{N}{2}}N!}}
\Phi^{n_k,n_j}_{s_{N-1,N}}(\mathbf{r}_{N-1},\mathbf{r}_{N})
		\lvert S,M\rangle,
	\end{align}
	the second set describe a case with two half-filled orbitals. They
	read
	\begin{align}\label{eq:TwoOrbital}
		\Psi^{n_k,n_j}_{S,M}(\mathbf{r}_1,\ldots,\mathbf{r}_{N}) = 
	\sum_{P_{\mathbf{r}_1\ldots\mathbf{r}_{N}}}
		c_{\mathbf{r}_{N-1},\mathbf{r}_{N}}
		\prod_{i}^{\frac{N}{2}-1}
		\frac{
			\Phi^i_{s_{2i-1,2i}}(\mathbf{r}_{2i-1},\mathbf{r}_{2i}) 
		}{\sqrt{2^{\frac{N}{2}}N!}}
	\Psi^{n_k,n_j}_{s_{N-1,N}}(\mathbf{r}_{N-1},\mathbf{r}_{N})
		\lvert S,M\rangle,
	\end{align}
\end{subequations}
where the sum runs over all permutations on the set of
coordinates $\mathbf{r}_1\ldots\mathbf{r}_{N}$ and the functions in
the summands are
\begin{subequations}
	\begin{align}
		\Phi^i_{s_{2i-1,2i}}(\mathbf{r}_{2i-1},\mathbf{r}_{2i})=
	\tfrac12 
		\big[ 
		\phi_i(\mathbf{r}_{2i-1})\phi_i(\mathbf{r}_{2i})
		-(-1)^{1+s_{2i-1,2i}}
	\phi_i(\mathbf{r}_{2i-1})\phi_i(\mathbf{r}_{2i})
		\big],
	\end{align}
	\begin{align}\label{eq:OneOrbitalSymmetrized_2}
		\Phi^{n_k,n_j}_{s_{N-1,N}}(\mathbf{r}_{N-1},\mathbf{r}_{N})=
		\tfrac{1}{\sqrt{2}} \big[ 
		\phi_{n_k}(\mathbf{r}_{N-1})\phi_{n_k}(\mathbf{r}_{N})
		-(-1)^{1+s_{N-1,N}}
	\phi_{n_j}(\mathbf{r}_{N-1})\phi_{n_j}(\mathbf{r}_{N}) \big]
	\end{align}
	and
	\begin{align}\label{eq:TwoOrbital_2}
		\Psi^{n_k,n_j}_{s_{N-1,N}}(\mathbf{r}_{N-1},\mathbf{r}_{N})=
	\tfrac{1}{\sqrt{2}}
		\big[ 
		\phi_{n_k}(\mathbf{r}_{N-1})\phi_{n_j}(\mathbf{r}_{N})
		-(-1)^{1+s_{N-1,N}}
	\phi_{n_j}(\mathbf{r}_{N-1})\phi_{n_k}(\mathbf{r}_{N})	
		\big].
	\end{align}
\end{subequations}
\end{widetext}
Moreover, for $i,j = 1,\ldots, N$ the permutation coefficients 
\[
c_{\mathbf{r}_{i},\mathbf{r}_{j}}= (-1)^{i+j+1},
\quad
c_{\mathbf{r}_{j},\mathbf{r}_{i}}=-(-1)^{1+s_{N-1,N}}(-1)^{i+j+1},
\]
account for the antisymmetry of triplet function in Eq. \eqref{eq:TwoOrbital_2}.
The spin part in the Eq. \eqref{eq:OneOrbitalSymmetrized} and 
\eqref{eq:TwoOrbital} is given by
\begin{equation*}
	\lvert S,M\rangle=\otimes^{\frac{N}{2}}_{i=1}
	\lvert s_{2i-1,2i},m_{2i-1,2i}\rangle.
\end{equation*}
Notice that $S$ and $M$ are not the total spin and magnetic quantum numbers
of the system. 
The latter represent the sets of all spin and magnetic quantum numbers for each 
pair of electrons, respectively.
Since we study a system with two effective spin-half centers one has
to bear in 
mind the following constraints
\begin{subequations}\label{eq:OrbitalSpinConstraints}
	\begin{equation}
		\sum^{N/2}_{i=1}
		\hat{\mathbf{s}}^2_{2i-1,2i}\lvert S,M\rangle
		= S(S+1)\lvert S,M\rangle,
	\end{equation}
	\begin{equation}
		\sum^{N/2}_{i=1}\hat{s}^z_{2i-1,2i}\lvert S,M\rangle
		=  M\lvert S,M\rangle,
	\end{equation}
\end{subequations}
with $S=0,1$ and $M=0,\pm1$.
We would like to point out that with respect to Eq. \eqref{eq:OrthogonalOrbitals} 
for all $k$ and $j$ the functions in Eq.
\eqref{eq:OneOrbitalSymmetrized} and \eqref{eq:TwoOrbital} are mutually
orthogonal.
Furthermore, they have to be used by having in mind that for $N=2$ 
their spatial parts reduce to Eq.
\eqref{eq:OneOrbitalSymmetrized_2} and 
\eqref{eq:TwoOrbital_2}, respectively.

\subsection{Key integrals}\label{sec:keyintegrals}	
Hamiltonian in Eq. \eqref{eq:CoulombHamilton} accounts for the kinetic
and repulsion energies of all electrons in the system.
As the $N/2$ lower orbitals are fully occupied
it is therefore sufficient to calculate the corresponding average 
energy value. Then we are left only with the operators
related to orbitals that are not fully occupied. 
Similar to the Hartree method \cite{hartree_1928,tsuneda_2014}
this procedure will demonstrate explicitly the contribution of
$(N/2+k)$-th orbitals
to the exchange processes.
Hence, with $N-2$ electrons, occupying all molecular orbitals lower in energy
than the $(N/2+k)$-th orbital, we have
\begin{align}\label{eq:LowOrbitalAverageEnergy}
	\hat{H}'=
	\prod_{i,j}^{\tfrac{N}{2}-1}
	\int\!\!\cdots\!\!\int
	\bar{\Phi}^j_{0}(\mathbf{r}_{2j-1},\mathbf{r}_{2j})
	\hat{H}
	\Phi^i_{0}(\mathbf{r}_{2i-1},\mathbf{r}_{2i})
	\mathrm{d}\mathbf{r}_{1}\ldots\mathrm{d}\mathbf{r}_{N-2}
	\nonumber \\
	\nonumber
\end{align}
Consequently, we distinguish four types of integrals related with the
processes of exchange and transfer of electrons. 
Considering the $(N/2+k)$-th orbital we obtain the integral
\begin{subequations}\label{eq:OrbitalIntergals}
	\begin{equation}\label{eq:OneOrbitIntegral}
		U_{n_k}=
		\iint
		\bar{\phi}_{n_k}(\mathbf{r}_{N-1})
		\bar{\phi}_{n_k}(\mathbf{r}_{N})
		\hat{H}'
		\phi_{n_k}(\mathbf{r}_{N-1})
		\phi_{n_k}(\mathbf{r}_{N})
		\mathrm{d}\mathbf{r}_{N-1}
		\mathrm{d}\mathbf{r}_{N},
	\end{equation}
	accounting for the kinetic and potential 
	energies of two electrons occupying the same orbital.
	The hopping integral
	\begin{equation}\label{eq:OrbitHoppingIntegral}
		t_{n_k}=
		\iint
		\bar{\phi}_{n_k}(\mathbf{r}_{N-1})
		\bar{\phi}_{n_k}(\mathbf{r}_{N})
		\hat{H}'
		\phi_{n_k}(\mathbf{r}_{N-1})
		\phi_{n_j}(\mathbf{r}_{N})
		\mathrm{d}\mathbf{r}_{N-1}
		\mathrm{d}\mathbf{r}_{N},
	\end{equation}
	associated with the transfer of an electron between two orbitals. 
	The integral
	\begin{equation}\label{eq:TwoOrbitIntegral}
		V_{n_kn_j}=
		\iint
		\bar{\phi}_{n_k}(\mathbf{r}_{N-1})
		\bar{\phi}_{n_j}(\mathbf{r}_{N})
		\hat{H}'
		\phi_{n_k}(\mathbf{r}_{N-1})
		\phi_{n_j}(\mathbf{r}_{N})
		\mathrm{d}\mathbf{r}_{N-1}
		\mathrm{d}\mathbf{r}_{N},
	\end{equation}
	representing the energy of two electrons occupying different orbitals 
	and the exchange integral
	\begin{equation}\label{eq:ExchamgeOrbitIntegral}
		D_{n_kn_j}=
		\iint
		\bar{\phi}_{n_k}(\mathbf{r}_{N-1})
		\bar{\phi}_{n_j}(\mathbf{r}_{N})
		\hat{H}'
		\phi_{n_j}(\mathbf{r}_{N-1})
		\phi_{n_k}(\mathbf{r}_{N})
		\mathrm{d}\mathbf{r}_{N-1}
		\mathrm{d}\mathbf{r}_{N},
	\end{equation}
\end{subequations}
associated with the energy of direct exchange of two 
electrons between orbitals $n_k$ and $n_j$.

The integrals in Eq. \eqref{eq:OneOrbitIntegral} and 
\eqref{eq:OrbitHoppingIntegral}
are nonzero only when the electron's spins are antiparallel, 
see Eq. \eqref{eq:OneOrbitalSymmetrized_2}. 
By analogy with a closed shell system the single orbital term 
describes a case with compactly occupied molecular
orbitals, i.e. a nonmagnetic molecule.
Therefore, Eq. \eqref{eq:OneOrbitIntegral}
and \eqref{eq:OrbitHoppingIntegral} favor antiferromagnetism
and refer to magnetic insulators.
If according to the Hund's rule a triplet state related to 
any two electrons occupying orbitals $N/2+k$ and $N/2+j$,
respectively,
is the ground state, then the integrals in Eq.
\eqref{eq:TwoOrbitIntegral} and \eqref{eq:ExchamgeOrbitIntegral} 
will determine the values of the transition energy.
Therefore, taking into account Eq. \eqref{eq:OrbitalSpinConstraints}
and \eqref{eq:OrbitalIntergals} we obtain the following relations

\begin{widetext}
\begin{align}\label{eq:OrbitalAverageEnergyTriplet}
	\int\!\!\cdots\!\!\int
	\bar{\Psi}^{n_k,n_j}_{1,M}(\mathbf{r}_1,\ldots,\mathbf{r}_{N})
	\hat{H}
	\Psi^{n_{k'},n_{j'}}_{1,M}(\mathbf{r}_1,\ldots,\mathbf{r}_{N})
	\mathrm{d}\mathbf{r}_{1}\ldots\mathrm{d}\mathbf{r}_{N}
	=\left( V_{n_kn_j}-D_{n_kn_j}\right)\delta_{kk'}\delta_{jj'}
	+\mathcal{O}
	\left( \tfrac{1}{\sqrt{2^{N-2}}N!}\right),
\end{align}
\begin{subequations}\label{eq:OrbitalAverageEnergySinglet}
	\begin{align}\label{eq:OrbitalAverageEnergySinglet_a}
		\int\!\!\cdots\!\!\int
	\bar{\Psi}^{n_k,n_j}_{0,0}(\mathbf{r}_1,\ldots,\mathbf{r}_{N})
		\hat{H}
		\Psi^{n_{k'},n_{j'}}_{0,0}(\mathbf{r}_1,\ldots,\mathbf{r}_{N})
		\mathrm{d}\mathbf{r}_{1}\ldots\mathrm{d}\mathbf{r}_{N}
	=\left( V_{n_kn_j}+D_{n_kn_j}\right)\delta_{kk'}\delta_{jj'}
		+\mathcal{O}
		\left( \tfrac{1}{\sqrt{2^{N-2}}N!}\right),
	\end{align}
	\begin{align}\label{eq:OrbitalAverageEnergySinglet_b}
		\int\!\!\cdots\!\!\int
		\bar{\Phi}^{n_k,n_j}_{0,0}(\mathbf{r}_1,\ldots,\mathbf{r}_{N})
		\hat{H}
		\Phi^{n_{k'},n_{j'}}_{0,0}(\mathbf{r}_1,\ldots,\mathbf{r}_{N})
		\mathrm{d}\mathbf{r}_{1}\ldots\mathrm{d}\mathbf{r}_{N}
	=\left( U_{n_kn_j}+D_{n_kn_j}\right)\delta_{kk'}\delta_{jj'}
		+\mathcal{O}
		\left( \tfrac{1}{\sqrt{2^{N-2}}N!}\right)
	\end{align}
	and
	\begin{align}\label{eq:OrbitalAverageEnergySinglet_c}
		\int\!\!\cdots\!\!\int
	\bar{\Psi}^{n_k,n_j}_{0,0}(\mathbf{r}_1,\ldots,\mathbf{r}_{N})
		\hat{H}
		\Phi^{n_{k'},n_{j'}}_{0,0}(\mathbf{r}_1,\ldots,\mathbf{r}_{N})
		\mathrm{d}\mathbf{r}_{1}\ldots\mathrm{d}\mathbf{r}_{N}
	=2 t_{n_kn_j} \delta_{kk'}\delta_{jj'} +\mathcal{O}
		\left( \tfrac{1}{\sqrt{2^{N-2}}N!}\right),
	\end{align}
\end{subequations}
\end{widetext}

where 
\[
t_{n_kn_j}=\frac{1}{2}\left( t_{n_k}+t_{n_j}\right),
\qquad
U_{n_kn_j}=\frac{1}{2}\left( U_{n_k}+U_{n_j}\right).
\]
Notice that the integrals in Eq. \eqref{eq:OrbitalAverageEnergySinglet}
vanish rapidly with $N$ for $k'\ne k$ and $j'\ne j$.
Therefore, in terms of matrices one can reduce the column 
and row number of the corresponding to Eq. \eqref{eq:OrbitalAverageEnergySinglet} 
matrix into a $(5\times5)$ and represent it by the sum $3\oplus2$.
The $(3\times3)$ matrix include the energy associated with the triplet 
group in Eq. \eqref{eq:OrbitalAverageEnergyTriplet} and the $(2\times2)$
one represent the energy related with the singlet states in Eq. 
\eqref{eq:OrbitalAverageEnergySinglet}.
Accordingly, the set of all eigenstates consists of the triplet group 
represented by the functions in Eq. \eqref{eq:OrbitalAverageEnergyTriplet} 
and the singlet group
\begin{subequations}\label{eq:SingletGroup}
	\begin{align}\label{eq:OmegaState}
		\Omega^{n_k,n_j}_{0,0}(\mathbf{r}_1,\ldots,\mathbf{r}_{N})=
	&	\frac{\sin\phi}{\sqrt{2}}
		\Phi^{n_{k},n_{j}}_{0,0}(\mathbf{r}_1,\ldots,\mathbf{r}_{N})
	\nonumber \\
	&	+
		\frac{\cos\phi}{\sqrt{2}}
		\Psi^{n_{k},n_{j}}_{0,0}(\mathbf{r}_1,\ldots,\mathbf{r}_{N}),
	\end{align}
	\begin{align}\label{eq:ThetaState}
		\Theta^{n_k,n_j}_{0,0}(\mathbf{r}_1,\ldots,\mathbf{r}_{N})=
	&	\frac{\cos\phi}{\sqrt{2}}
		\Phi^{n_{k},n_{j}}_{0,0}(\mathbf{r}_1,\ldots,\mathbf{r}_{N})
	\nonumber \\
	&	-
		\frac{\sin\phi}{\sqrt{2}}
		\Psi^{n_{k},n_{j}}_{0,0}(\mathbf{r}_1,\ldots,\mathbf{r}_{N}),
	\end{align}
\end{subequations}
where 
{\small 
	\[
	\phi=\arctan
	\left( 
	\frac{4t_{n_kn_j}}{U_{n_kn_j}-V_{n_kn_j}+
		\sqrt{16t^2_{n_kn_j}+\left(U_{n_kn_j}-V_{n_kn_j}\right)^2}}
	\right).
	\]
}

From now on we consider only the state in Eq. \eqref{eq:OmegaState}
since it correspond to the lower energy value for all $t_{n_kn_j}$.

\section{MULTIPLE EXCHANGE PATHWAYS}\label{sec:pathways}
For isolated diatomic and triatomic magnetic units
the exchange pathways are unique.
In general, the Goodenough-Kanamori-Anderson rules 
\cite{anderson_1950,goodenough_1955,kanamori_1959,anderson_1959} holds and the magnetic spectra 
are usually explained in terms of the Heisenberg model
\cite{heisenberg_1926}.
With respect to the nature of ligands in some magnetic compounds,
anisotropic spin Hamiltonians 
are valuable for studying the magnetic properties.
Further, in the case of mono or diatomic intermediate bridges
in periodic latices the competition 
between kinetic energy and Coulomb repulsion can be adequately studied
within the framework of the Hubbard model \cite{hubbard_1963,hubbard_1964}.
However, in complex molecular magnets, the exchange
process between two effective magnetic centers involves a number of
intermediate nonmagnetic ions. 
Thus, if two such centers are connected by more than one intricate bridge, 
see for example the molecular magnet Ni$_4$MO$_{12}$ \cite{muller_2000},
it is possible to have more than
one energetically favorable distribution of unpaired valence electrons.
Accordingly, multiple independent 
magnetic excitations that do not arise due to the anisotropy related with
spin-orbit coupling nor to the existence of electronic bands, but
rather results from the activation of different exchange bridges, will
emerge. A sign for the absence of a unique exchange bridge can be the
broadened excitation peaks in the observed magnetic spectrum and the enhanced
response of the molecular magnet to an external magnetic field.
In particular since the values of the overlap integrals depend on the angles between 
coupled ions the effect of applied magnetic field or changes in 
temperature may cause 
variations in the energy of the considered molecular orbitals altering 
the distribution of valence electrons and hence the values 
of the transition energy.

\subsection{The effective Hamiltonian}
In order to simplify all further expressions we change the notations
by setting $\tau=(n_k,n_j)$ as a general index denoting
both the half filled molecular orbitals and the number of valence 
electrons. Therefore, as the functions in Eq. \eqref{eq:DimerStateFunctions} 
correspond to a certain distribution of electrons,
$\tau$ will indicates all existing exchange bridges.
Then, for $\tau'\ne\tau$ one has different number of electrons $N'\ne N$.
Further, as the number of electrons is related to the number of
all spin pairs one has $S'\ne S$ and $M'\ne M$.
Then, in order to address the aforementioned assumptions correctly 
we label the number of valence electrons $N$ and
the sets of spin and magnetic quantum numbers $S$ and $M$,
according to the corresponding bridge obtaining
$N_{\tau}$, $S_\tau$ and $M_\tau$, respectively.
Hamiltonian in Eq. \eqref{eq:CoulombHamilton} depends on 
the number of valence electrons and coupled ions.
Henceforward we will be denoting the later by 
$\hat{H}_\tau$. 

Since the unpaired electrons can be distributed over 
any of the considered bridges, the generic state functions
will account for all probabilities. Thus, setting 
$\mathbf{r}_\tau=\{\mathbf{r}_1,\ldots,\mathbf{r}_{N_\tau}\}$ and 
using the states in Eq. \eqref{eq:OneOrbitalSymmetrized},
\eqref{eq:TwoOrbital} and \eqref{eq:OmegaState},
within the current notations we obtain the triplet states
\begin{equation}\label{eq:AllPathsStatesTriplet}
	\varPsi_{1,m}(\mathbf{r}_\tau,\ldots,\mathbf{r}_{\tau'})
	=\sum_{\tau}c^\tau_{1} 
	\Psi^{\tau}_{1,M_\tau}(\mathbf{r}_{\tau}),
	\qquad
	\forall \ S_\tau=1
\end{equation}
and the singlet state
\begin{equation}\label{eq:AllPathsStatesSinglet}
	\varPsi_{0,0}(\mathbf{r}_\tau,\ldots,\mathbf{r}_{\tau'})
	=\sum_{\tau}c^\tau_{0} 
	\Omega^{\tau}_{0,0}(\mathbf{r}_{\tau}),
	\qquad
	\forall \ S_\tau=0,
\end{equation}
where for all $\tau$ the quantities $s=S_\tau$, $m=M_\tau$ 
are the dimer effective spin and magnetic quantum numbers.
The last functions are orthonormal and the coefficients
$c^\tau_{S_\tau}\in\mathbb{R}$ depend on the spin quantum 
numbers since for the
singlet state, $S_\tau=0$, the two electrons are
closer to each other than in the case of triplet states, $S_\tau=1$.

We would like to point out that according to the direct sum of 
the spin subspaces of the exchange pathways one has
$
\langle S_{\tau^{}},M_{\tau^{}}|S_{\tau'},M_{\tau'}\rangle
=
\delta_{\tau^{}\tau'}
$.
Notice also that for all $\tau$ the constraints in Eq. 
\eqref{eq:OrbitalSpinConstraints} are always satisfied.

Taking into consideration the integrals in Eq.
\eqref{eq:OrbitalAverageEnergyTriplet} and
\eqref{eq:OrbitalAverageEnergySinglet} 
for all $\tau$ the energy expectation values read
\begin{equation}\label{eq:OnePathEnergyExpectation}
\begin{array}{l}
	\displaystyle
	E^\tau_{1,M_\tau}=
	\frac{\int
		\bar{\Psi}^\tau_{1,M_\tau}(\mathbf{r}_{\tau})
		\hat{H}_{\tau}
		\Psi^\tau_{1,M_\tau}(\mathbf{r}_{\tau})
		\mathrm{d}\mathbf{r}_{\tau}
	}{
		\int
		\bar{\Psi}^\tau_{1,M_\tau}(\mathbf{r}_{\tau})
		\Psi^\tau_{1,M_\tau}(\mathbf{r}_{\tau})
		\mathrm{d}\mathbf{r}_{\tau}
	},
	\\ [0.7cm]
	\displaystyle
	E^\tau_{0,0}=
	\frac{\int
		\bar{\Omega}^\tau_{0,0}(\mathbf{r}_{\tau})
		\hat{H}_{\tau}
		\Omega^\tau_{0,0}(\mathbf{r}_{\tau})
		\mathrm{d}\mathbf{r}_{\tau}
	}{
		\int
		\bar{\Omega}^\tau_{0,0}(\mathbf{r}_{\tau})
		\Omega^\tau_{0,0}(\mathbf{r}_{\tau})
		\mathrm{d}\mathbf{r}_{\tau}
	},
\end{array}
\end{equation}
where 
$
\mathrm{d}\mathbf{r}_{\tau}=
\mathrm{d}\mathbf{r}_{1}\ldots\mathrm{d}\mathbf{r}_{N_\tau}.
$
In terms of the integrals in Eq.
\eqref{eq:OrbitalIntergals}, the energy values in Eq.
\eqref{eq:OnePathEnergyExpectation} are given by 
\begin{subequations}\label{eq:EffSpectrum}
	\begin{align}\label{eq:EffTriplet}
		E^\tau_{1,M_\tau}=V_\tau-D_\tau,
		\qquad
		M_\tau=0,\pm1,
	\end{align}
	\begin{align}\label{eq:EffSinglet}
		E^\tau_{0,0}= D_\tau+\frac{U_\tau+V_\tau}{2}-
		\sqrt{4t^2_\tau+\frac{\left(U_\tau-V_\tau\right)^2}{4}},
	\end{align}
\end{subequations}	
respectively.
Therefore,
the expectation values of the Hamiltonian in Eq. \eqref{eq:CoulombHamilton}
obtained from states written in Eq. \eqref{eq:AllPathsStatesTriplet} and 
\eqref{eq:AllPathsStatesSinglet} can then be written as
\begin{equation}\label{eq:AllPathExpectation}
	E_{s,m}=\sum_{\tau} \big|c^\tau_{S_\tau}\big|^2
	E^{\tau}_{S_\tau,M_\tau}.
\end{equation}
Notice that the Hamiltonians related with the different bridging structures 
commute, $[\hat{H}_{\tau},\hat{H}_{\tau'}]=0$.
The last energy values 
represent the spectrum of an effective Hamiltonian satisfying
\begin{equation*}
	\hat{H}_{eff}\lvert s,m \rangle = E_{s,m}\lvert s,m \rangle,
\end{equation*}
where the states $\lvert s,m \rangle$ are eigenstates of the 
effective dimer total spin operator 
$\hat{\mathbf{s}}=(\hat{s}^x,\hat{s}^y,\hat{s}^z)$, with
\begin{equation}\label{eq:TotalSpin}
	\hat{\mathbf{s}}^2\lvert s,m\rangle= s(s+1)\lvert s,m\rangle,
	\qquad
	\hat{s}^z\lvert s,m\rangle= m\lvert s,m\rangle.
\end{equation}
With respect to the spin quantum number $s$ the Hamiltonian is $(2s+1)$-fold
degenerate and hence the considered system allows only one transition, 
$\big|E_{1,m}-E_{0,0}\big|$. 
On the other hand, as the energy values in Eq.
\eqref{eq:AllPathExpectation} depend on the probability coefficients in Eq.
\eqref{eq:AllPathsStatesTriplet} and \eqref{eq:AllPathsStatesSinglet},
multiple transitions related to the spatial part of the
state functions are allowed.
However, relying on such assumptions, one has to take into account 
that the energy conservation law do not allow the simultaneous
existence of more than one excitation. 
Thereby, for the transition energy we obtain
\begin{equation}\label{eq:TransitionEnergy}
	\left|\Delta E\right|=
	\sum_{\tau} 
	\left| \big|c^\tau_{1}\big|^2 E^{\tau}_{1,M_\tau}
	-\big|c^\tau_{0}\big|^2 E^{\tau}_{0,0}\right|.
\end{equation}
Although the coefficients in Eq. \eqref{eq:TransitionEnergy} can be 
represented by analytical functions one can observe a number of different
values for $\Delta E$ related with a discrete spectrum. 
This follows from the independence of all possible exchange 
pathways and the difference of the electron's behavior at triplet and singlet states.
As a consequence, for $s=0$ and at certain conditions the electrons 
could be localized only on one of the exchange bridges.
In contrary, for the triplet state both electrons could be distributed over 
all bridges.

\subsection{The simplest case}\label{ssec:simple}
Assuming a molecule with unique exchange bridge of one or two
intermediate atoms the sum in Eq. \eqref{eq:TransitionEnergy} is reduced
to a single term with $\big|c^\tau_{1}\big|^2=1$ and
$\big|c^\tau_{0}\big|^2=1$.
Consequently, suggesting an antiferromagnetic ground state and taking
into account Eq. \eqref{eq:EffSpectrum}, for 
the transition energy in Eq. \eqref{eq:TransitionEnergy} we have
\begin{align}\label{eq:TransitionEnergySimple}
	\Delta E=
	-2D_\tau-
	\frac{1}{2}\left(U_\tau+V_\tau\right)+
	\frac{1}{2}
	\sqrt{16t^2_\tau+
		\left( U_\tau-V_\tau\right)^2}.
\end{align}
Within the selected ground state
the value of integral in Eq. \eqref{eq:TwoOrbitIntegral} could be considered as 
rather smaller than 
the single orbital integral in Eq. \eqref{eq:OneOrbitIntegral} and one can
use of the inequality $U_\tau\gg V_\tau$.
Furthermore, since the intermediate and magnetic ions are of different
kind, their mutual orientation renders
the value of corresponding overlap integral negligible.
This implies that $U_\tau>t_\tau$ and $t_\tau>D_\tau$,
see Eq. \eqref{eq:OrbitalIntergals}. 
Perturbing over $t_\tau/U_\tau$ and taking into
account only the first two terms from the series, we obtain
\begin{equation}\label{eq:SuperExchangeOrbit}
	\Delta E=\frac{4t^2_\tau}{U_\tau}-2D_\tau.
\end{equation}	
For some compounds the direct exchange term $D_\tau$ is 
assumed negligible.
The ground state is antiferromagnetic and the system behaves as a magnetic
insulator, since $t_\tau$ and $U_\tau$ favor antifferomagnetism, see Eq.
\eqref{eq:OneOrbitalSymmetrized} and \eqref{eq:OneOrbitalSymmetrized_2}.
A different approach describing localized electrons by 
using Wannier functions \cite{marzari_2012} and leading to 
analogous conclusions are introduced in Ref. \cite{anderson_1963}. 

\section{THE SPIN HAMILTONIAN}\label{sec:spinmodel}
Within the spin space, the magnetic excitation energy is 
associated only with molecular orbitals that are not fully occupied.
Therefore, the total spin of the chemical complex is effectively taken into 
account by considering the number of unpaired electrons. Then, all magnetic
features are interpreted in terms of either effective or fictitious spins
of magnetic centers usually representing transition metal ions.

Operating with a conventional bilinear spin Hamiltonian, one 
won't be able to
associate each of the excitations in Eq. \eqref{eq:TransitionEnergy} 
within a single singlet-triplet transition. 
Therefore, we pursue a different approach to account for the 
features arising from the existence of multiple exchange pathways.

Let $i=1,2$ indicate both effective magnetic centers in the considered  spin dimer
and $\hat{\mathbf{s}}_i=(\hat{s}^\alpha_i)_{\alpha\in\mathbb{K}}$ are their
corresponding spin operators with
\[
\hat{\mathbf{s}}^2_i\lvert s_i,m_i\rangle= s_i(s_i+1)\lvert s_i,m_i\rangle,
\qquad
\hat{s}^z_i\lvert s_i,m_i\rangle= m_i\lvert s_i,m_i\rangle,
\]
where $s_i$ and $m_i$ are the respective spin and magnetic quantum numbers. 
Notice that in fact the last spin operators account for the spins of both unpaired electrons.
Further, let 
$\hat{\mathbf{s}}=\hat{\mathbf{s}}_1+\hat{\mathbf{s}}_2$ be the total spin
operator written in Eq. \eqref{eq:TotalSpin}.	  	 
In order to obtain an energy spectrum consistent with the 
transitions in Eq. \eqref{eq:TransitionEnergy} we propose the following Hamiltonian
\begin{equation}\label{eq:SpinHamiltonian}
	\hat{\mathcal{H}}= J 
	\left( 
	\hat{\mathbf{s}}_1\cdot\hat{\boldsymbol{\sigma}}_2
	+
	\hat{\mathbf{s}}_2\cdot\hat{\boldsymbol{\sigma}}_1
	\right)
	-
	g\mu_{\mathrm{B}}\hat{s}^zB, 
\end{equation}
where $J$ is an effective constant, $g$ is the isotropic $g$-tensor,
$\mu_{\mathrm{B}}$ is the Bohr magneton, $B$ represents the externally 
applied magnetic field and the operator
$\hat{\boldsymbol{\sigma}}_i =
(\hat{\sigma}^x_i, \hat{\sigma}^y_i, \hat{\sigma}^z_i)$ 
effectively accounts for the differences in valence electron's distribution
with respect to the $i$-th magnetic center.
We would like to point out that the atomic orbitals of
selected magnetic centers are considered as quenched and in
accordance with the proposed superposition in Eq. \eqref{eq:MOrbitals}
the $g$-tensor does not alter. 

Since the excitation energy is related with the transition between
singlet and triplet states of the total dimer spin space one has to account
for the total $\sigma$-operator defined by
\begin{equation}\label{eq:Sigma}
	\hat{\sigma}^\alpha \lvert s,m \rangle
	=
	a_{s,m} \hat{s}^{\alpha_{\phantom{j}}} 
	\lvert s,m \rangle,
\end{equation}
where $a_{s,m} \in \mathbb{R}$.
As the spins of both magnetic centers are paired their 
relevant $\sigma$-operators share the coefficient in Eq. \eqref{eq:Sigma} 
and govern the transformations
\begin{equation}\label{eq:Sigma_k}
	\hat{\sigma}^{\alpha}_i \lvert s,m \rangle
	=
	a_{s,m} \hat{s}^{\alpha}_i 
	\lvert s,m \rangle,
	\qquad
	i=1,2.
\end{equation}
The rising and lowering $\sigma$-operators corresponding to Eq.
\eqref{eq:Sigma} satisfy
\begin{equation}\label{eq:Sigma_LR}
	\hat{\sigma}^{\pm} \lvert s,m \rangle	
	=
	a_{s,m} \hat{s}^{\pm} 
	\lvert s,m \rangle,
\end{equation}
where $\hat{s}^{\pm}$ are the total rising and lowering spin operators.
Thereby, taking into consideration Eq. \eqref{eq:TotalSpin}, \eqref{eq:Sigma}, 
\eqref{eq:Sigma_LR} and the following three cases $m=s$,
$-s<m<s$ and $m=-s$, where $s\ne0$ for the eigenvalues of total sigma 
square operator we have
\begin{subequations}\label{eq:SigmaSquareEigenvalue}
	\begin{equation}\label{eq:SigmaSquareEigenvalue_m=s}
		a^2_{s,s} s^2+a^{}_{s,s} a^{}_{s,s-1} s,
	\end{equation}
	\begin{align}\label{eq:SigmaSquareEigenvalue_m}
	&	\tfrac12 a_{s,m} \left[a_{s,m+1} + 
		a_{s,m-1}\right]s(s+1)+
		a^2_{s,m} m^2
	\nonumber \\ 
	-& \tfrac12 a_{s,m} m 
		\left[a_{s,m+1}(m+1) 
		+ a_{s,m-1} (m-1)\right],
	\end{align}
	\begin{equation}\label{eq:SigmaSquareEigenvalue_m=-s}
		a^2_{s,-s} s^2+a^{}_{s,-s} a^{}_{s,1-s} s.
	\end{equation}
\end{subequations}
The eigenvalues of $\sigma^z$ can be obtained directly from Eq. \eqref{eq:Sigma}.

All excitations are a result of the transition between 
singlet and triplet states, see Eq. \eqref{eq:TransitionEnergy}, 
where the total spin and magnetic quantum numbers 
are always preserved, see also Eq.	\eqref{eq:OrbitalSpinConstraints}.
Therefore, we imply the following constraints
\begin{subequations}\label{eq:ConstraintSigma}
	\begin{equation}\label{eq:SigmaZ}
		\hat{\sigma}^z\lvert s,m\rangle=
		h_s m\lvert s,m\rangle,
	\end{equation}
	\begin{equation}\label{eq:SigmaSquare}
		\hat{\boldsymbol{\sigma}}^2 \lvert s,m\rangle =
		h^2_s s(s+1) \lvert s,m \rangle,
	\end{equation}
\end{subequations}
where the parameters $h_s$ account for the changes of
electrons distributions in the molecule and thus the variation 
of all coefficients in Eq. \eqref{eq:MOrbitals}, 
\eqref{eq:AllPathsStatesTriplet} and \eqref{eq:AllPathsStatesSinglet}
due to the action of $B$. For $B=0$ and all $s$ we have $h_s=1$.

Within the last constraints one distinguish three cases:
\begin{enumerate}
	\item[(1)] $s \ne 0$ and $m \ne 0$, then
	\begin{equation}\label{eq:asm}
		a_{s,m\pm1}=a_{s,m}=h_s.
	\end{equation}
	As a consequence, for $B=0$, the $\sigma$-operator transforms the
	spin eigenstates as the spin operator does
	and the Hamiltonian in Eq. \eqref{eq:SpinHamiltonian} coincide with its Heisenberg 
	counterpart.
	
	\item[(2)] $s \ne 0$ and $m=0$, then the coefficients in Eq. \eqref{eq:Sigma} 
	are restricted only
	by the Eq. \eqref{eq:SigmaSquare}. As a result, from Eq. \eqref{eq:SigmaZ}, 
	\eqref{eq:SigmaSquareEigenvalue_m} and \eqref{eq:SigmaSquare} one obtains
	\begin{equation}\label{eq:as0}
		a_{s,\pm1}=a_{s,0}=\pm h_s.
	\end{equation}
	It is important to remark that the minus sign is an intrinsic feature of 
	the sigma operators rather than related to the effectively included spatial part 
	of the state functions in Eq. \eqref{eq:OneOrbitalSymmetrized} and \eqref{eq:TwoOrbital}.
	
	\item[(3)] $s = 0$, then $a_{0,0}$ remains unconstrained 
	and there exist a set of parameters $c_{n}\in\mathbb{R}$ 
	such that
	\begin{equation}\label{eq:a00}
		a_{0,0} \in \{h_0 c_n\}_{n\in\mathbb{N}}.
	\end{equation} 
	In general, $c_n$ are represented as a function of
	the integrals in Eq. \eqref{eq:OrbitalIntergals} and therefore depend
	on the number of valence electrons and ions, taking part in the
	exchange process.
\end{enumerate}

\section{TWO EXCITATIONS}\label{sec:excitations}
Consider two exchange bridges, $\tau=a,b$, see FIG. \ref{fig:Dimer}. 
Let for $s=1$ the 
probability both electrons to be distributed over a given
exchange bridges be given by the coefficients $c^a_{1}$ and
$c^b_{1}$, respectively. Thus, if for a singlet state and at 
temperature $T_1$ both electrons are localized on the 
bridge $a$ and at temperature $T_2$, $b$ is the more energetically
favorable bridge, then from Eq. \eqref{eq:TransitionEnergy} we distinguish
two transitions
\begin{subequations}\label{eq:TwoTransitions}
	\begin{equation}
		\Delta E_1= 
		\big| c^{a}_{1}\big|^2 E^{a}_{1,M_a}+
		\big| c^b_{1}\big|^2 E^{b}_{1,M_b}
		-\big| c^a_{0}\big|^2 E^{a}_{0,0}
	\end{equation}
	and
	\begin{equation}
		\Delta E_2= 
		\big| c^a_{1}\big|^2 E^{a}_{1,M_a}+
		\big| c^b_{1}\big|^2 E^{b}_{1,M_b}
		-\big| c^b_{0}\big|^2 E^{b}_{0,0},
	\end{equation}
\end{subequations}
respectively.

\begin{figure}[ht!]
	\begin{tikzpicture}[scale=1.5]
	\draw[->,dotted](-0.6,0)--(3.8,0);
	\node[circle
	,line width=0pt
	,shading=ball
	,ball color=blue
	,shading angle=-80
	,scale=1.3 
	](Ni1) at (0.05,0.1) {};
	
	\node[circle
	,line width=0pt
	,shading=ball
	,ball color=blue
	,shading angle=-80 
	,scale=1.3
	](Ni2) at (3,0) {};
	
	\Edge[color=blue!50!white](Ni1)(Ni2); 
	\Edge[color=red!70!white](Ni2)(Ni1);	
	\node[ 
	,circle
	,dashed
	,line width=0pt
	,shading=ball
	,ball color=white
	,shading angle=-80 
	,scale=0.5
	](Z0) at (0.4,0.25) {};
	
	\node[ 
	,circle
	,dashed
	,line width=0pt
	,shading=ball
	,ball color=white
	,shading angle=-80 
	,scale=0.5
	](Z00) at (2.6,0.25) {};
	
	\node[ 
	,circle
	,dashed
	,line width=0pt
	,shading=ball
	,ball color=white
	,shading angle=-80 
	,scale=0.5
	](Z1) at (1.18,0.55) {};
	
	\node[ 
	,circle
	,dashed
	,line width=0pt
	,shading=ball
	,ball color=white
	,shading angle=-80 
	,scale=0.5
	](Z3) at (1.85,0.55) {};
	
	\node[ 
	,circle
	,dashed
	,line width=0pt
	,shading=ball
	,ball color=white
	,shading angle=-80 
	,scale=0.5
	](Z5) at (1.85,0.45) {};
	
	\node[ 
	,circle
	,dashed
	,line width=0pt
	,shading=ball
	,ball color=white
	,shading angle=-80 
	,scale=0.5
	](Z7) at (1.17,0.45) {};
	
	\node[ 
	,circle
	,dashed
	,line width=0pt
	,shading=ball
	,ball color=orange
	,shading angle=-80 
	,scale=0.8
	](X1) at (0.75,0.5) {};
	
	\node[ 
	,circle
	,dashed
	,line width=0pt
	,shading=ball
	,ball color=orange
	,shading angle=-80 
	,scale=0.8
	](X3) at (2.25,0.5) {};
	
	\node[ 
	,circle
	,dashed
	,line width=0pt
	,shading=ball
	,ball color=gray
	,shading angle=-80 
	,scale=1
	](Y1) at (1.5,0.7) {};
	\node[ 
	,circle
	,dashed
	,line width=0pt
	,shading=ball
	,ball color=gray
	,shading angle=-80 
	,scale=1
	](Y3) at (1.5,0.3) {};
	\node[ 
	,circle
	,dashed
	,line width=0pt
	,shading=ball
	,ball color=white
	,shading angle=-80 
	,scale=0.5
	](Z2) at (1.18,-0.55) {};
	
	\node[ 
	,circle
	,dashed
	,line width=0pt
	,shading=ball
	,ball color=white
	,shading angle=-80 
	,scale=0.5
	](Z4) at (1.85,-0.55) {};
	
	\node[ 
	,circle
	,dashed
	,line width=0pt
	,shading=ball
	,ball color=white
	,shading angle=-80 
	,scale=0.5
	](Z6) at (1.85,-0.65) {};
	
	\node[ 
	,circle
	,dashed
	,line width=0pt
	,shading=ball
	,ball color=white
	,shading angle=-80 
	,scale=0.5
	](Z8) at (1.17,-0.65) {};
	
	\node[ 
	circle
	,dashed
	,line width=0pt
	,shading=ball
	,ball color=orange
	,shading angle=-80 
	,scale=0.8
	](X2) at (0.9,-0.55) {};
	
	\node[
	circle
	,dashed
	,line width=0pt
	,shading=ball
	,ball color=orange
	,shading angle=-80 
	,scale=0.8
	](X4) at (2.1,-0.55) {};
	
	\node[ 
	,circle
	,dashed
	,line width=0pt
	,shading=ball
	,ball color=gray
	,shading angle=-80 
	,scale=1
	](Y2) at (1.5,-0.45) {};
	\node[ 
	,circle
	,dashed
	,line width=0pt
	,shading=ball
	,ball color=gray
	,shading angle=-80 
	,scale=1
	](Y4) at (1.5,-0.75) {};
	
	\draw[](Ni1)--(Z0)--(X1)--(Z1)--(Y1)--(Z3)--(X3)--(Z00)--(Ni2);
	\draw[](Ni1)--(Z0)--(X1)--(Z7)--(Y3)--(Z5)--(X3)--(Z00)--(Ni2);

	\draw[](Ni1)--(X2)--(Z2)--(Y2)--(Z4)--(X4)--(Ni2);
	\draw[](Ni1)--(X2)--(Z8)--(Y4)--(Z6)--(X4)--(Ni2);
	
	\draw[dashed](Ni1)--(-0.3,0.3);
	\draw[dashed](Ni1)--(-0.3,-0.3);
	\draw[dashed](Ni2)--(3.3,0.3);
	\draw[dashed](Ni2)--(3.3,-0.3);
	
	\draw[dashed](X2)--(0.6,-0.6);	
	\draw[dashed](X4)--(2.4,-0.6);
	
	\node[]() at (3.6,-0.2){{\small $x$}}; 
	\node[]() at (1,0.9){{\small $a$}}; 
	\node[]() at (2.1,-0.9){{\small $b$}}; 
	\end{tikzpicture}
	\caption{Model of dimeric magnetic molecule with non symmetric exchange bridges. 
		The effective magnetic centers are colored in blue.
		The upper and lower bridges are labeled by the letter $a$ and 
		$b$, respectively.
		The light blue arrow depicts the bridge related with the low temperature transition,
		the light red arrow shows the high temperature analogue, 
		see Eq. \eqref{eq:TwoTransitions}.}
	\label{fig:Dimer}
\end{figure}
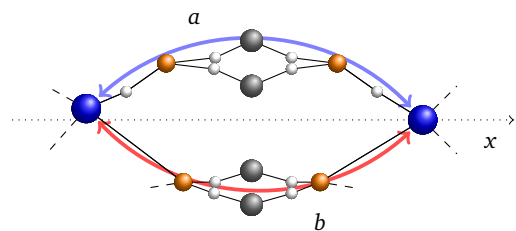

\begin{figure*}[ht!]
	\begin{center}
		\begin{tikzpicture}
		\def\a{1.5}	
		\def\l{1.1}	
		\begin{scope}
		\draw[->,black, line width=1 pt]
		(0,-0.2) to (0,3.5);
		\node[rotate=90]() at (-0.3,3){ [meV]};		
		\draw [black, line width=\l pt, domain=0.2:4] plot (\x, 0); 
		\node[]() at (-0.4,-0.1){{\small $\mathcal{E}^{(2)}_{0,0}$}};
		
		\draw [black, line width=\l pt, domain=0.2:4] plot (\x, 0.25); 
		\node[]() at (-0.4,0.4){{\small $\mathcal{E}^{(1)}_{0,0}$}};
		
		\draw [black, line width=\l pt, domain=0.2:4] plot (\x, 1); 
		\node[]() at (-0.4,1.1){{\small $\mathcal{E}^{(2)}_{1,0}$}};
		
		\draw [black, line width=\l pt, domain=0.2:4] plot (\x, 2); 
		\node[]() at (-0.4,2){{\small $\mathcal{E}^{(1)}_{1,m}$}};
		\draw[<->, >=stealth,line width=\a pt, blue] (1,0) to (1,2); 
		\node[]() at (1.45,0.5){{\small $|\Delta E_1|$}};
		\draw[<->, >=stealth,line width=\a pt, red] (2,0.25) to (2,2); 
		\node[]() at (2.5,0.6){{\small $|\Delta E_2|$}};
		
		\node[draw,rectangle]() at (2,3.3){{\small $\mathrm{B}=0$}};
		
		\end{scope}
		\begin{scope}{xshift=4cm}
		\draw[dashed,line width=\l pt] (4,0) to (4,3.5);
		\draw[dashed,line width=\l pt] (5,0) to (5,3.5);
		\draw[->,line width=\l pt] (4,0)--(4.5,0.25*0.2)--(5,0.25*0.2);
		\draw[->,line width=\l pt] (4,0.25)--(4.5,0.25*1.2)--(5,0.25*1.2);
		\draw[->,line width=\l pt] (4,1)--(4.5,1.3)--(5,1.3);
		\draw[->,line width=\l pt] (4,2)--(4.5,2*1.3)--(5,2*1.3);
		\node[]() at (4.5,3){{\small $m=0$}};
		\end{scope}
		\begin{scope}{xshift=5cm}
		\draw [black, line width=\l pt, domain=5.1:9] plot (\x, 0.25*0.2); 
		\draw [black, line width=\l pt, domain=5.1:9] plot (\x, 0.25*1.2); 
		\draw [black, line width=\l pt, domain=5.1:9] plot (\x, 1*1.3); 
		\draw [black, line width=\l pt, domain=5.1:9] plot (\x, 2*1.3); 
		\node[draw,rectangle]() at (7,3.3){{\small $\mathrm{B}\ne0$, $m=0$}};
		\draw[<->, >=stealth,line width=\a pt, blue] (6,0.25*0.2) to (6,2*1.3); 
		\node[]() at (6.45,0.6){{\small $|\Delta E^{*}_1|$}};
		\draw[<->, >=stealth,line width=\a pt, red] (7,0.25*1.2) to (7,2*1.3); 
		\node[]() at (7.5,0.75){{\small $|\Delta E^{*}_2|$}};
		\end{scope}
		\end{tikzpicture}
		\caption{ Energy spectrum of the spin-half dimer obtained 
			using the Hamiltonian in Eq. \eqref{eq:SpinHamiltonian}.
			On the left hand side the spectrum for $B=0$ is depicted.
			The right hand side spectrum shows the energy levels shifting
			due to the applied external magnetic field, where for brevity 
			only the levels associated with the nonmagnetic states are
			illustrated. The blue and red arrows show both
			transitions, corresponding to the bridges on FIG. \ref{fig:Dimer}.
		}
		\label{fig:EnSpectrum}
	\end{center}
\end{figure*}
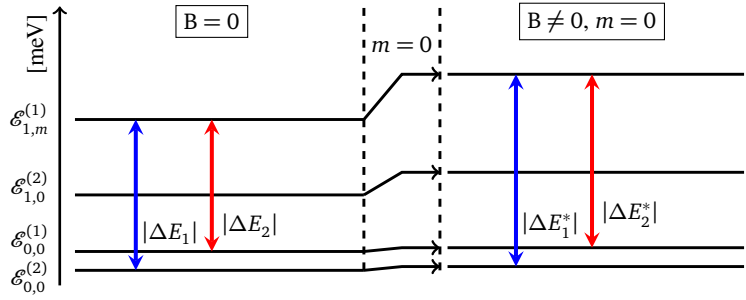

\subsection{Energy spectrum in the absence of an external magnetic field}
\label{ssec:spectparameters}
As the total spin commute with the Hamiltonian in Eq. \eqref{eq:SpinHamiltonian},
see Eq. \eqref{eq:TotalSpin} and \eqref{eq:Sigma_k}, for $B=0$ we may write 
\begin{equation}\label{eq:SpinHamiltonianSpectrum}
\hat{\mathcal{H}}\lvert s,m\rangle=
2Ja_{s,m}
\left( 
\hat{\mathbf{s}}_1\cdot\hat{\mathbf{s}}_2
\right) 
\lvert s,m\rangle.
\end{equation}
In order to clarify the advantage of Eq. \eqref{eq:SpinHamiltonianSpectrum}
we assume $T_1$$<$$T_2$ and focus on the corresponding transition energies in Eq.
\eqref{eq:TwoTransitions}.
Hence, with respect to the quantum numbers $s$ and $m$, from Eq.
\eqref{eq:SpinHamiltonianSpectrum} we distinguish four energy values
\begin{equation}\label{eq:Energy_sm}
\tfrac12Ja_{1,+1},
\quad
\tfrac12Ja_{1,0},
\quad
\tfrac12Ja_{1,-1},
\quad
-\tfrac32Ja_{0,0},
\end{equation}
where the first three quantities, from left to right, correspond to the triplet group
and the last one is associated with the singlet state.
Nevertheless, without accounting for the discrete values of the coefficients 
in Eq. \eqref{eq:Energy_sm} the excitation energy 
$\big|\Delta E_2\big|$ will remain unexplained.
Therefore
taking into account Eq. \eqref{eq:asm}, \eqref{eq:as0}, \eqref{eq:a00}
and suggesting that $\big|\Delta E_2\big|<\big|\Delta E_1\big|$ from Eq. \eqref{eq:Energy_sm}
we obtain an energy spectrum consisting of four levels, see FIG. 
\ref{fig:EnSpectrum}.

Two related with the triplet and two with the singlet states.
For $m=0,\pm1$ and $n=1,2$ the set of all energy values read 
\begin{equation}\label{eq:EnergySpectrum}
	\mathcal{E}^{(1)}_{1,m}=\tfrac12J,
	\quad
	\mathcal{E}^{(2)}_{1,0}=-\tfrac12J,
	\quad
	\mathcal{E}^{(n)}_{0,0}=-\tfrac32Jc_n,
\end{equation}	
According to the energy conservation law and Eq. \eqref{eq:TransitionEnergy}
we have 
\[
\Delta E_1=\mathcal{E}^{(2)}_{0,0}-\mathcal{E}^{(1)}_{1,m},
\quad
\Delta E_2=\mathcal{E}^{(1)}_{0,0}-\mathcal{E}^{(1)}_{1,m}.
\]
Hence, for the model parameters we obtain
\begin{equation}\label{eq:Parameters}
	J=-\frac{\Delta E_1}{2},
	\qquad
	c_1=\frac{4}{3}\frac{\Delta E_2}{\Delta E_1}-\frac13,
	\qquad
	c_2=1.
\end{equation}
Depending on the type of exchange, we may obtain different
values for the quantities $J$ and $c_1$,
see Eq. \eqref{eq:TransitionEnergySimple} and \eqref{eq:SuperExchangeOrbit}.
The value of $c_1$ can be determined from spectroscopic
measurements. 

Extracting all coefficients $a_{s,m}$ from Eq. \eqref{eq:Energy_sm} one can 
clearly distinguish the spectrum obtained from the Heisenberg model and
compare it with the spectrum in Eq. \eqref{eq:EnergySpectrum}. 
Within the framework of conventional spin bilinear Hamiltonians 
one assume the functions in Eq. \eqref{eq:OneOrbitalSymmetrized},
\eqref{eq:TwoOrbital} and the coefficients in Eq. \eqref{eq:TransitionEnergy} 
as unique and works only with $J$ in Eq. \eqref{eq:Parameters}.
In such case, the exchange coupling is given by the relation
\begin{equation*}
	J=
	D_\tau+
	\frac{1}{4}\left(U_\tau+V_\tau\right)+
	\frac{1}{4}
	\sqrt{16t^2_\tau-
		\left( U_\tau-V_\tau\right)^2}
\end{equation*}
and to account for the
additional excitation a procedure of searching for different spin interaction 
terms may be started. Nevertheless, we would like to
emphasize that the origin
of the transition $\Delta E_2$ cannot be explained by the inclusion of
anisotropy or higher order spin interaction terms.

\subsection{Energy spectrum with applied external magnetic field}	
\label{ssec:magparameters}
The applied external magnetic field  
affects the probability distribution of valence electrons in 
molecular magnets. Such phenomena can be quantitatively studied by 
including the parameters $h_s$ from Eq. \eqref{eq:ConstraintSigma} 
that take into account the 
variation of all coefficients in Eq. \eqref{eq:MOrbitals},
\eqref{eq:AllPathsStatesTriplet} and \eqref{eq:AllPathsStatesSinglet}.
Therefore, considering Eq. \eqref{eq:TotalSpin}, 
\eqref{eq:SpinHamiltonianSpectrum} and \eqref{eq:EnergySpectrum}, for
$B\ne0$ we get the magnetic part of the energy spectrum
\begin{equation}\label{eq:EnergySpectrumMagnetic_B}
	\mathcal{E}_{1,+1}=\tfrac12Jh_1-g\mu_{\mathrm{B}}B,
	\qquad
	\mathcal{E}_{1,-1}=\tfrac12Jh_1+g\mu_{\mathrm{B}}B,
\end{equation}
and for $n=1,2$ the nonmagnetic part
\begin{equation}\label{eq:EnergySpectrum_B}
	\mathcal{E}^{(1)}_{1,0}=\tfrac12Jh_1,
	\quad
	\mathcal{E}^{(2)}_{1,0}=-\tfrac12Jh_1,
		\quad
	\mathcal{E}^{(n)}_{0,0}=-\tfrac32Jh_0c_n,
\end{equation}	
where we omitted the superscript in Eq. \eqref{eq:EnergySpectrumMagnetic_B}
since it has no contribution.
Calculating the energy level shifting, $\Delta E_1\to\Delta E^{*}_1$, 
and $\Delta E_2\to\Delta E^{*}_2$, we have to
take into consideration only the energy levels associated with
the nonmagnetic states, see FIG. \ref{fig:EnSpectrum}.
Thus, accounting for the energy
conservation law, from Eq. \eqref{eq:EnergySpectrum_B} we have
\begin{equation}\label{eq:TransitionEnergy_B}
	\Delta E^{*}_1=\mathcal{E}^{(2)}_{0,0}-\mathcal{E}^{(1)}_{1,0},
	\qquad
	\Delta E^{*}_2=\mathcal{E}^{(1)}_{0,0}-\mathcal{E}^{(1)}_{1,0}.
\end{equation}
Using the relations in Eq. \eqref{eq:TransitionEnergy_B} together with the parameters in Eq.
\eqref{eq:Parameters} for the field parameters we obtain 
\begin{equation}\label{eq:FieldParameters}
\begin{array}{l}
\displaystyle
	h_0=\frac{
	\Delta E^{*}_2-\Delta E^{*}_1
	}{
	\Delta E_2-\Delta E_1
	}
	\\ [0.5cm]
\displaystyle	
	h_1=2\frac{\Delta E^{*}_2+\Delta E^{*}_1}{\Delta E_1}
	-h_0
	\frac{2\Delta E_2+\Delta E_1}{\Delta E_1}.
\end{array}
\end{equation}
The values of $h_s$ can be fixed form the magnetization and 
low-temperature susceptibility measurements. Therefore, as the 
absolute temperature alter the values of both parameters in Eq.
\eqref{eq:FieldParameters} in order to calculate the contribution 
of $B$ correctly one has 
to perform measurements only at very low temperatures.

\section{CONCLUSION}\label{sec:conclusion}	
We studied the role of bridging complexes in the exchange processes
and evaluated their contribution 
within the framework of a spin-half dimer molecular magnet.
To account for the influence of the intermediate structure
we assume that the overall structure consists of more that one 
favorable exchange pathway.
Within the framework of such assumptions none of the conventional spin 
models provide an appropriate energy spectrum. Therefore, to address all 
relevant features we proposed a formalism based on an adequate 
bilinear spin Hamiltonian, Eq. \eqref{eq:SpinHamiltonian}.
It is important to emphasize that
with respect to a certain representation 
the $\sigma$-operators are not unique.

Although the exchange Hamiltonian in Eq. \eqref{eq:SpinHamiltonian} 
is an approximate model, 
it may describe reasonably well the magnetism in real
compounds. 	
Recently \cite{georgiev_2018} we demonstrated the application 
of Eq. \eqref{eq:SpinHamiltonian} in larger spin clusters.
We analyzed the inelastic neutron scattering
spectrum of the trimeric compounds A$_3$Cu$_3$(PO$_4$)$_4$,
A=(Ca, Sr, Pb)
and Nickel tetramer spin cluster Ni$_4$Mo$_{12}$ and
we obtained results consistent with the available spectroscopic
measurements \cite{matsuda_magnetic_2005,podlesnyak_magnetic_2007,nehrkorn_inelastic_2010,furrer_magnetic_PRB_2010}.

In conclusion, we would like to point out that the existence of more than
one transition at the same temperature would have been possible
assuming more than two unpaired electrons and more than two magnetic centers. 
The probability to observe the aforementioned features increases
with the size of the magnetic cluster.
In the real compounds a distortion in structure's symmetry would have to be 
taken into account for the observation of a peculiar
magnetic spectrum.
Nevertheless, in compounds with only two or three distinct ions
and periodic
structure the discussed features cannot be observed, see Eq. \eqref{eq:TransitionEnergySimple}. 
Therefore, the application of the proposed method remains restricted to a 
specific variety of spin clusters. 
For example, clusters in which the electrons are not localized around
a certain ion and 
on the other hand are not a part of conduction band.

\section*{ACKNOWLEDGMENTS}
This work was supported by the Bulgarian National Science Fund under
contract DN/08/18.

%


\begin{thebibliography}{54}%
	\makeatletter
	\providecommand \@ifxundefined [1]{%
		\@ifx{#1\undefined}
	}%
	\providecommand \@ifnum [1]{%
		\ifnum #1\expandafter \@firstoftwo
		\else \expandafter \@secondoftwo
		\fi
	}%
	\providecommand \@ifx [1]{%
		\ifx #1\expandafter \@firstoftwo
		\else \expandafter \@secondoftwo
		\fi
	}%
	\providecommand \natexlab [1]{#1}%
	\providecommand \enquote  [1]{``#1''}%
	\providecommand \bibnamefont  [1]{#1}%
	\providecommand \bibfnamefont [1]{#1}%
	\providecommand \citenamefont [1]{#1}%
	\providecommand \href@noop [0]{\@secondoftwo}%
	\providecommand \href [0]{\begingroup \@sanitize@url \@href}%
	\providecommand \@href[1]{\@@startlink{#1}\@@href}%
	\providecommand \@@href[1]{\endgroup#1\@@endlink}%
	\providecommand \@sanitize@url [0]{\catcode `\\12\catcode `\$12\catcode
		`\&12\catcode `\#12\catcode `\^12\catcode `\_12\catcode `\%12\relax}%
	\providecommand \@@startlink[1]{}%
	\providecommand \@@endlink[0]{}%
	\providecommand \url  [0]{\begingroup\@sanitize@url \@url }%
	\providecommand \@url [1]{\endgroup\@href {#1}{\urlprefix }}%
	\providecommand \urlprefix  [0]{URL }%
	\providecommand \Eprint [0]{\href }%
	\providecommand \doibase [0]{http://dx.doi.org/}%
	\providecommand \selectlanguage [0]{\@gobble}%
	\providecommand \bibinfo  [0]{\@secondoftwo}%
	\providecommand \bibfield  [0]{\@secondoftwo}%
	\providecommand \translation [1]{[#1]}%
	\providecommand \BibitemOpen [0]{}%
	\providecommand \bibitemStop [0]{}%
	\providecommand \bibitemNoStop [0]{.\EOS\space}%
	\providecommand \EOS [0]{\spacefactor3000\relax}%
	\providecommand \BibitemShut  [1]{\csname bibitem#1\endcsname}%
	\let\auto@bib@innerbib\@empty
	\bibitem [{\citenamefont {Hay}\ \emph {et~al.}(1975)\citenamefont {Hay},
		\citenamefont {Thibeault},\ and\ \citenamefont {Hoffmann}}]{hay_1975}%
	\BibitemOpen
	\bibfield  {author} {\bibinfo {author} {\bibfnamefont {P.~Jeffrey}\
			\bibnamefont {Hay}}, \bibinfo {author} {\bibfnamefont {Jack~C.}\ \bibnamefont
			{Thibeault}}, \ and\ \bibinfo {author} {\bibfnamefont {Roald}\ \bibnamefont
			{Hoffmann}},\ }\bibfield  {title} {\enquote {\bibinfo {title} {Orbital
				interactions in metal dimer complexes},}\ }\href {\doibase
		10.1021/ja00850a018} {\bibfield  {journal} {\bibinfo  {journal} {J. Am. Chem.
				Soc.}\ }\textbf {\bibinfo {volume} {97}},\ \bibinfo {pages} {4884} (\bibinfo
		{year} {1975})}\BibitemShut {NoStop}%
	\bibitem [{\citenamefont {Felthouse}\ \emph {et~al.}(1977)\citenamefont
		{Felthouse}, \citenamefont {Laskowski},\ and\ \citenamefont
		{Hendrickson}}]{felthouse_1977}%
	\BibitemOpen
	\bibfield  {author} {\bibinfo {author} {\bibfnamefont {Timothy~R.}\
			\bibnamefont {Felthouse}}, \bibinfo {author} {\bibfnamefont {Edward~J.}\
			\bibnamefont {Laskowski}}, \ and\ \bibinfo {author} {\bibfnamefont
			{David~N.}\ \bibnamefont {Hendrickson}},\ }\bibfield  {title} {\enquote
		{\bibinfo {title} {Magnetic exchange interactions in transition metal dimers.
				10. {Structural} and magnetic characterization of oxalate-bridged,
				bis(1,1,4,7,7-pentaethyldiethylene triamine)oxalatodicopper tetraphenylborate
				and related dimers. {Effects} of nonbridging ligands and counterions on
				exchange interactions},}\ }\href {\doibase 10.1021/ic50171a023} {\bibfield
		{journal} {\bibinfo  {journal} {Inorg. Chem.}\ }\textbf {\bibinfo {volume}
			{16}},\ \bibinfo {pages} {1077} (\bibinfo {year} {1977})}\BibitemShut
	{NoStop}%
	\bibitem [{\citenamefont {Lor\"{o}sch}\ \emph {et~al.}(1987)\citenamefont
		{Lor\"{o}sch}, \citenamefont {Quotschalla},\ and\ \citenamefont
		{Haase}}]{lorosch_1987}%
	\BibitemOpen
	\bibfield  {author} {\bibinfo {author} {\bibfnamefont {J\"{u}rgen}\
			\bibnamefont {Lor\"{o}sch}}, \bibinfo {author} {\bibfnamefont {Udo}\
			\bibnamefont {Quotschalla}}, \ and\ \bibinfo {author} {\bibfnamefont
			{Wolfgang}\ \bibnamefont {Haase}},\ }\bibfield  {title} {\enquote {\bibinfo
			{title} {Magneto-structural dependencies for asymmetrically bridged
				{Cu}({II}) dimers},}\ }\href {\doibase 10.1016/S0020-1693(00)96030-4}
	{\bibfield  {journal} {\bibinfo  {journal} {Inorganica Chim. Acta}\ }\textbf
		{\bibinfo {volume} {131}},\ \bibinfo {pages} {229} (\bibinfo {year}
		{1987})}\BibitemShut {NoStop}%
	\bibitem [{\citenamefont {Gehring}\ \emph {et~al.}(1993)\citenamefont
		{Gehring}, \citenamefont {Fleischhauer}, \citenamefont {Paulus},\ and\
		\citenamefont {Haase}}]{gehring_1993}%
	\BibitemOpen
	\bibfield  {author} {\bibinfo {author} {\bibfnamefont {Stefan}\ \bibnamefont
			{Gehring}}, \bibinfo {author} {\bibfnamefont {Peter}\ \bibnamefont
			{Fleischhauer}}, \bibinfo {author} {\bibfnamefont {Helmut}\ \bibnamefont
			{Paulus}}, \ and\ \bibinfo {author} {\bibfnamefont {Wolfgang}\ \bibnamefont
			{Haase}},\ }\bibfield  {title} {\enquote {\bibinfo {title} {Ferromagnetic
				exchange coupling and magneto-structural correlations in mixed-bridged
				trinuclear copper({II}) complexes. {Magnetic} data and theoretical
				investigations and crystal structures of two angled {CuII}3 complexes},}\
	}\href {\doibase 10.1021/ic00053a009} {\bibfield  {journal} {\bibinfo
			{journal} {Inorg. Chem.}\ }\textbf {\bibinfo {volume} {32}},\ \bibinfo
		{pages} {54} (\bibinfo {year} {1993})}\BibitemShut {NoStop}%
	\bibitem [{\citenamefont {Astheimer}\ and\ \citenamefont
		{Haase}(1986)}]{astheimer_1986}%
	\BibitemOpen
	\bibfield  {author} {\bibinfo {author} {\bibfnamefont {Harald}\ \bibnamefont
			{Astheimer}}\ and\ \bibinfo {author} {\bibfnamefont {Wolfgang}\ \bibnamefont
			{Haase}},\ }\bibfield  {title} {\enquote {\bibinfo {title} {Direct
				theoretical \textit{a}\textit{b}
				\textit{i}\textit{n}\textit{i}\textit{t}\textit{i}\textit{o} calculations in
				exchange coupled copper ({II}) dimers: {Influence} of structural and chemical
				parameters in modeled copper dimers},}\ }\href {\doibase 10.1063/1.451232}
	{\bibfield  {journal} {\bibinfo  {journal} {J. Chem. Phys.}\ }\textbf
		{\bibinfo {volume} {85}},\ \bibinfo {pages} {1427} (\bibinfo {year}
		{1986})}\BibitemShut {NoStop}%
	\bibitem [{\citenamefont {Fraser}\ \emph {et~al.}(2017)\citenamefont {Fraser},
		\citenamefont {Nichol}, \citenamefont {Velmurugan}, \citenamefont
		{Rajaraman},\ and\ \citenamefont {Brechin}}]{fraser_2017}%
	\BibitemOpen
	\bibfield  {author} {\bibinfo {author} {\bibfnamefont {Hector W.~L.}\
			\bibnamefont {Fraser}}, \bibinfo {author} {\bibfnamefont {Gary~S.}\
			\bibnamefont {Nichol}}, \bibinfo {author} {\bibfnamefont {Gunasekaran}\
			\bibnamefont {Velmurugan}}, \bibinfo {author} {\bibfnamefont {Gopalan}\
			\bibnamefont {Rajaraman}}, \ and\ \bibinfo {author} {\bibfnamefont {Euan~K.}\
			\bibnamefont {Brechin}},\ }\bibfield  {title} {\enquote {\bibinfo {title}
			{Magneto-structural correlations in a family of di-alkoxo bridged chromium
				dimers},}\ }\href {\doibase 10.1039/C7DT01197K} {\bibfield  {journal}
		{\bibinfo  {journal} {Dalton Trans.}\ }\textbf {\bibinfo {volume} {46}},\
		\bibinfo {pages} {7159} (\bibinfo {year} {2017})}\BibitemShut {NoStop}%
	\bibitem [{\citenamefont {Charlot}\ \emph {et~al.}(1986)\citenamefont
		{Charlot}, \citenamefont {Kahn}, \citenamefont {Chaillet},\ and\
		\citenamefont {Larrieu}}]{charlot_1986}%
	\BibitemOpen
	\bibfield  {author} {\bibinfo {author} {\bibfnamefont {Marie~France.}\
			\bibnamefont {Charlot}}, \bibinfo {author} {\bibfnamefont {Olivier.}\
			\bibnamefont {Kahn}}, \bibinfo {author} {\bibfnamefont {Max.}\ \bibnamefont
			{Chaillet}}, \ and\ \bibinfo {author} {\bibfnamefont {Christiane.}\
			\bibnamefont {Larrieu}},\ }\bibfield  {title} {\enquote {\bibinfo {title}
			{Interaction between copper({II}) ions through the azido bridge: concept of
				spin polarization and ab initio calculations on model systems},}\ }\href
	{\doibase 10.1021/ja00270a014} {\bibfield  {journal} {\bibinfo  {journal} {J.
				Am. Chem. Soc.}\ }\textbf {\bibinfo {volume} {108}},\ \bibinfo {pages} {2574}
		(\bibinfo {year} {1986})}\BibitemShut {NoStop}%
	\bibitem [{\citenamefont {Aebersold}\ \emph {et~al.}(1998)\citenamefont
		{Aebersold}, \citenamefont {Gillon}, \citenamefont {Plantevin}, \citenamefont
		{Pardi}, \citenamefont {Kahn}, \citenamefont {Bergerat}, \citenamefont {von
			Seggern}, \citenamefont {Tuczek}, \citenamefont {\"{O}hrstr\"{o}m},
		\citenamefont {Grand},\ and\ \citenamefont
		{Leli\`{e}vre-Berna}}]{aebersold_1998}%
	\BibitemOpen
	\bibfield  {author} {\bibinfo {author} {\bibfnamefont {Michael~A.}\
			\bibnamefont {Aebersold}}, \bibinfo {author} {\bibfnamefont {B\'{e}atrice}\
			\bibnamefont {Gillon}}, \bibinfo {author} {\bibfnamefont {Olivier}\
			\bibnamefont {Plantevin}}, \bibinfo {author} {\bibfnamefont {Luca}\
			\bibnamefont {Pardi}}, \bibinfo {author} {\bibfnamefont {Olivier}\
			\bibnamefont {Kahn}}, \bibinfo {author} {\bibfnamefont {Pierre}\ \bibnamefont
			{Bergerat}}, \bibinfo {author} {\bibfnamefont {Ingo}\ \bibnamefont {von
				Seggern}}, \bibinfo {author} {\bibfnamefont {Felix}\ \bibnamefont {Tuczek}},
		\bibinfo {author} {\bibfnamefont {Lars}\ \bibnamefont {\"{O}hrstr\"{o}m}},
		\bibinfo {author} {\bibfnamefont {Andr\'{e}}\ \bibnamefont {Grand}}, \ and\
		\bibinfo {author} {\bibfnamefont {E.}~\bibnamefont {Leli\`{e}vre-Berna}},\
	}\bibfield  {title} {\enquote {\bibinfo {title} {Spin {Density} {Maps} in the
				{Triplet} {Ground} {State} of [{Cu}$_{\textrm{2}}$
				(\textit{t}-{Bupy})$_{\textrm{4}}$({N}$_{\textrm{3}}$)$_{\textrm{2}}$]({ClO}$_{\textrm{4}}$)
				$_{\textrm{2}}$(\textit{t}-{Bupy}=\textit{p}-\textit{tert}-butylpyridine):
				{A} {Polarized} {Neutron} {Diffraction} {Study}},}\ }\href {\doibase
		10.1021/ja9739603} {\bibfield  {journal} {\bibinfo  {journal} {J. Am. Chem.
				Soc.}\ }\textbf {\bibinfo {volume} {120}},\ \bibinfo {pages} {5238} (\bibinfo
		{year} {1998})}\BibitemShut {NoStop}%
	\bibitem [{\citenamefont {Ribas}\ \emph {et~al.}(1993)\citenamefont {Ribas},
		\citenamefont {Monfort}, \citenamefont {Diaz}, \citenamefont {Bastos},\ and\
		\citenamefont {Solans}}]{ribas_1993}%
	\BibitemOpen
	\bibfield  {author} {\bibinfo {author} {\bibfnamefont {Joan}\ \bibnamefont
			{Ribas}}, \bibinfo {author} {\bibfnamefont {Montserrat}\ \bibnamefont
			{Monfort}}, \bibinfo {author} {\bibfnamefont {Carmen}\ \bibnamefont {Diaz}},
		\bibinfo {author} {\bibfnamefont {Carles}\ \bibnamefont {Bastos}}, \ and\
		\bibinfo {author} {\bibfnamefont {Xavier}\ \bibnamefont {Solans}},\
	}\bibfield  {title} {\enquote {\bibinfo {title} {New antiferromagnetic
				dinuclear complexes of nickel({II}) with two azides as bridging ligands.
				{Magneto}-structural correlations},}\ }\href {\doibase 10.1021/ic00068a029}
	{\bibfield  {journal} {\bibinfo  {journal} {Inorg. Chem.}\ }\textbf {\bibinfo
			{volume} {32}},\ \bibinfo {pages} {3557} (\bibinfo {year}
		{1993})}\BibitemShut {NoStop}%
	\bibitem [{\citenamefont {Ribas}\ \emph {et~al.}(1999)\citenamefont {Ribas},
		\citenamefont {Escuer}, \citenamefont {Monfort}, \citenamefont {Vicente},
		\citenamefont {Cort\'{e}s}, \citenamefont {Lezama},\ and\ \citenamefont
		{Rojo}}]{ribas_1999}%
	\BibitemOpen
	\bibfield  {author} {\bibinfo {author} {\bibfnamefont {Joan}\ \bibnamefont
			{Ribas}}, \bibinfo {author} {\bibfnamefont {Albert}\ \bibnamefont {Escuer}},
		\bibinfo {author} {\bibfnamefont {Montserrat}\ \bibnamefont {Monfort}},
		\bibinfo {author} {\bibfnamefont {Ramon}\ \bibnamefont {Vicente}}, \bibinfo
		{author} {\bibfnamefont {Roberto}\ \bibnamefont {Cort\'{e}s}}, \bibinfo
		{author} {\bibfnamefont {Luis}\ \bibnamefont {Lezama}}, \ and\ \bibinfo
		{author} {\bibfnamefont {Te\'{o}filo}\ \bibnamefont {Rojo}},\ }\bibfield
	{title} {\enquote {\bibinfo {title} {Polynuclear {NiII} and {MnII} azido
				bridging complexes. {Structural} trends and magnetic behavior},}\ }\href
	{\doibase 10.1016/S0010-8545(99)00051-X} {\bibfield  {journal} {\bibinfo
			{journal} {Coord. Chem. Rev.}\ }\textbf {\bibinfo {volume} {193-195}},\
		\bibinfo {pages} {1027} (\bibinfo {year} {1999})}\BibitemShut {NoStop}%
	\bibitem [{\citenamefont {Sadhu}\ \emph {et~al.}(2017)\citenamefont {Sadhu},
		\citenamefont {Mathoniere}, \citenamefont {Patil},\ and\ \citenamefont
		{Kumar}}]{sadhu_2017}%
	\BibitemOpen
	\bibfield  {author} {\bibinfo {author} {\bibfnamefont {Mehul~H.}\
			\bibnamefont {Sadhu}}, \bibinfo {author} {\bibfnamefont {Corine}\
			\bibnamefont {Mathoniere}}, \bibinfo {author} {\bibfnamefont {Yogesh~P.}\
			\bibnamefont {Patil}}, \ and\ \bibinfo {author} {\bibfnamefont {Sujit~Baran}\
			\bibnamefont {Kumar}},\ }\bibfield  {title} {\enquote {\bibinfo {title}
			{Binuclear copper({II}) complexes with {N}3s-coordinate tripodal ligand and
				mixed azide-carboxylate bridges: {Synthesis}, crystal structures and magnetic
				properties},}\ }\href {\doibase 10.1016/j.poly.2016.11.036} {\bibfield
		{journal} {\bibinfo  {journal} {Polyhedron}\ }\textbf {\bibinfo {volume}
			{122}},\ \bibinfo {pages} {210} (\bibinfo {year} {2017})}\BibitemShut
	{NoStop}%
	\bibitem [{\citenamefont {Zhao}\ \emph {et~al.}(2017)\citenamefont {Zhao},
		\citenamefont {Deng}, \citenamefont {Zhou}, \citenamefont {Shao},
		\citenamefont {Wu}, \citenamefont {Wei},\ and\ \citenamefont
		{Wang}}]{zhao_2017}%
	\BibitemOpen
	\bibfield  {author} {\bibinfo {author} {\bibfnamefont {Xin-Hua}\ \bibnamefont
			{Zhao}}, \bibinfo {author} {\bibfnamefont {Lin-Dan}\ \bibnamefont {Deng}},
		\bibinfo {author} {\bibfnamefont {Yan}\ \bibnamefont {Zhou}}, \bibinfo
		{author} {\bibfnamefont {Dong}\ \bibnamefont {Shao}}, \bibinfo {author}
		{\bibfnamefont {Dong-Qing}\ \bibnamefont {Wu}}, \bibinfo {author}
		{\bibfnamefont {Xiao-Qin}\ \bibnamefont {Wei}}, \ and\ \bibinfo {author}
		{\bibfnamefont {Xin-Yi}\ \bibnamefont {Wang}},\ }\bibfield  {title} {\enquote
		{\bibinfo {title} {Slow {Magnetic} {Relaxation} in {One}-{Dimensional}
				{Azido}-{Bridged} {Co}$^{\textrm{{ii}}}$ {Complexes}},}\ }\href {\doibase
		10.1021/acs.inorgchem.7b00736} {\bibfield  {journal} {\bibinfo  {journal}
			{Inorg. Chem.}\ }\textbf {\bibinfo {volume} {56}},\ \bibinfo {pages} {8058}
		(\bibinfo {year} {2017})}\BibitemShut {NoStop}%
	\bibitem [{\citenamefont {Angaridis}\ \emph {et~al.}(2005)\citenamefont
		{Angaridis}, \citenamefont {Kampf},\ and\ \citenamefont
		{Pecoraro}}]{angaridis_2005}%
	\BibitemOpen
	\bibfield  {author} {\bibinfo {author} {\bibfnamefont {Panagiotis}\
			\bibnamefont {Angaridis}}, \bibinfo {author} {\bibfnamefont {Jeff~W.}\
			\bibnamefont {Kampf}}, \ and\ \bibinfo {author} {\bibfnamefont {Vincent~L.}\
			\bibnamefont {Pecoraro}},\ }\bibfield  {title} {\enquote {\bibinfo {title}
			{Multinuclear {Fe}({III}) {Complexes} with {Polydentate} {Ligands} of the
				{Family} of {Dicarboxyimidazoles}: {Nuclearity}- and {Topology}-{Controlled}
				{Syntheses} and {Magneto}-{Structural} {Correlations}},}\ }\href {\doibase
		10.1021/ic0481879} {\bibfield  {journal} {\bibinfo  {journal} {Inorg. Chem.}\
		}\textbf {\bibinfo {volume} {44}},\ \bibinfo {pages} {3626} (\bibinfo {year}
		{2005})}\BibitemShut {NoStop}%
	\bibitem [{\citenamefont {Mekuimemba}\ \emph {et~al.}(2018)\citenamefont
		{Mekuimemba}, \citenamefont {Conan}, \citenamefont {Mota}, \citenamefont
		{Palacios}, \citenamefont {Colacio},\ and\ \citenamefont
		{Triki}}]{mekuimemba_2018}%
	\BibitemOpen
	\bibfield  {author} {\bibinfo {author} {\bibfnamefont {Cle~Donacier}\
			\bibnamefont {Mekuimemba}}, \bibinfo {author} {\bibfnamefont {Fran\c{c}oise}\
			\bibnamefont {Conan}}, \bibinfo {author} {\bibfnamefont {Antonio~J.}\
			\bibnamefont {Mota}}, \bibinfo {author} {\bibfnamefont {Maria~A.}\
			\bibnamefont {Palacios}}, \bibinfo {author} {\bibfnamefont {Enrique}\
			\bibnamefont {Colacio}}, \ and\ \bibinfo {author} {\bibfnamefont {Smail}\
			\bibnamefont {Triki}},\ }\bibfield  {title} {\enquote {\bibinfo {title} {On
				the {Magnetic} {Coupling} and {Spin} {Crossover} {Behavior} in {Complexes}
				{Containing} the {Head}-to-{Tail}
				[{Fe}$^{\textrm{{ii}}}$$_{\textrm{2}}$($\mu$-{SCN})$_{\textrm{2}}$] {Bridging}
				{Unit}: {A} {Magnetostructural} {Experimental} and {Theoretical} {Study}},}\
	}\href {\doibase 10.1021/acs.inorgchem.7b03082} {\bibfield  {journal}
		{\bibinfo  {journal} {Inorg. Chem.}\ }\textbf {\bibinfo {volume} {57}},\
		\bibinfo {pages} {2184} (\bibinfo {year} {2018})}\BibitemShut {NoStop}%
	\bibitem [{\citenamefont {Gregoli}\ \emph {et~al.}(2009)\citenamefont
		{Gregoli}, \citenamefont {Danieli}, \citenamefont {Barra}, \citenamefont
		{Neugebauer}, \citenamefont {Pellegrino}, \citenamefont {Poneti},
		\citenamefont {Sessoli},\ and\ \citenamefont {Cornia}}]{gregoli_2009}%
	\BibitemOpen
	\bibfield  {author} {\bibinfo {author} {\bibfnamefont {Luisa}\ \bibnamefont
			{Gregoli}}, \bibinfo {author} {\bibfnamefont {Chiara}\ \bibnamefont
			{Danieli}}, \bibinfo {author} {\bibfnamefont {Anne-Laure}\ \bibnamefont
			{Barra}}, \bibinfo {author} {\bibfnamefont {Petr}\ \bibnamefont
			{Neugebauer}}, \bibinfo {author} {\bibfnamefont {Giovanna}\ \bibnamefont
			{Pellegrino}}, \bibinfo {author} {\bibfnamefont {Giordano}\ \bibnamefont
			{Poneti}}, \bibinfo {author} {\bibfnamefont {Roberta}\ \bibnamefont
			{Sessoli}}, \ and\ \bibinfo {author} {\bibfnamefont {Andrea}\ \bibnamefont
			{Cornia}},\ }\bibfield  {title} {\enquote {\bibinfo {title}
			{Magnetostructural {Correlations} in {Tetrairon}({III}) {Single}-{Molecule}
				{Magnets}},}\ }\href {\doibase 10.1002/chem.200900483} {\bibfield  {journal}
		{\bibinfo  {journal} {Chem. Eur. J.}\ }\textbf {\bibinfo {volume} {15}},\
		\bibinfo {pages} {6456} (\bibinfo {year} {2009})}\BibitemShut {NoStop}%
	\bibitem [{\citenamefont {Viennois}\ \emph {et~al.}(2010)\citenamefont
		{Viennois}, \citenamefont {Giannini}, \citenamefont {van~der Marel},\ and\
		\citenamefont {\v{C}ern\'{y}}}]{viennois_2010}%
	\BibitemOpen
	\bibfield  {author} {\bibinfo {author} {\bibfnamefont {R.}~\bibnamefont
			{Viennois}}, \bibinfo {author} {\bibfnamefont {E.}~\bibnamefont {Giannini}},
		\bibinfo {author} {\bibfnamefont {D.}~\bibnamefont {van~der Marel}}, \ and\
		\bibinfo {author} {\bibfnamefont {R.}~\bibnamefont {\v{C}ern\'{y}}},\
	}\bibfield  {title} {\enquote {\bibinfo {title} {Effect of {Fe} excess on
				structural, magnetic and superconducting properties of single-crystalline
				{Fe}$_{1+x}${Te}$_{1-y}${Se}$_y$},}\ }\href {\doibase
		10.1016/j.jssc.2010.01.024} {\bibfield  {journal} {\bibinfo  {journal} {J.
				Solid State Chem.}\ }\textbf {\bibinfo {volume} {183}},\ \bibinfo {pages}
		{769} (\bibinfo {year} {2010})}\BibitemShut {NoStop}%
	\bibitem [{\citenamefont {Loose}\ \emph {et~al.}(2008)\citenamefont {Loose},
		\citenamefont {Ruiz}, \citenamefont {Kersting},\ and\ \citenamefont
		{Kortus}}]{loose_2008}%
	\BibitemOpen
	\bibfield  {author} {\bibinfo {author} {\bibfnamefont {Claudia}\ \bibnamefont
			{Loose}}, \bibinfo {author} {\bibfnamefont {Eliseo}\ \bibnamefont {Ruiz}},
		\bibinfo {author} {\bibfnamefont {Berthold}\ \bibnamefont {Kersting}}, \ and\
		\bibinfo {author} {\bibfnamefont {Jens}\ \bibnamefont {Kortus}},\ }\bibfield
	{title} {\enquote {\bibinfo {title} {Magnetic exchange interaction in triply
				bridged dinickel({II}) complexes},}\ }\href {\doibase
		10.1016/j.cplett.2007.12.035} {\bibfield  {journal} {\bibinfo  {journal}
			{Chemical Physics Letters}\ }\textbf {\bibinfo {volume} {452}},\ \bibinfo
		{pages} {38--43} (\bibinfo {year} {2008})}\BibitemShut {NoStop}%
	\bibitem [{\citenamefont {Panja}\ \emph {et~al.}(2017)\citenamefont {Panja},
		\citenamefont {Jana}, \citenamefont {Adak}, \citenamefont {Brand\~{a}o},
		\citenamefont {Dlh\'{a}\v{n}}, \citenamefont {Titi\v{s}},\ and\ \citenamefont
		{Bo\v{c}a}}]{panja_2017}%
	\BibitemOpen
	\bibfield  {author} {\bibinfo {author} {\bibfnamefont {Anangamohan}\
			\bibnamefont {Panja}}, \bibinfo {author} {\bibfnamefont {Narayan~Ch.}\
			\bibnamefont {Jana}}, \bibinfo {author} {\bibfnamefont {Sarmistha}\
			\bibnamefont {Adak}}, \bibinfo {author} {\bibfnamefont {Paula}\ \bibnamefont
			{Brand\~{a}o}}, \bibinfo {author} {\bibfnamefont {Lubor}\ \bibnamefont
			{Dlh\'{a}\v{n}}}, \bibinfo {author} {\bibfnamefont {J\'{a}n}\ \bibnamefont
			{Titi\v{s}}}, \ and\ \bibinfo {author} {\bibfnamefont {Roman}\ \bibnamefont
			{Bo\v{c}a}},\ }\bibfield  {title} {\enquote {\bibinfo {title} {The structure
				and magnetism of mono- and di-nuclear {Ni}({II}) complexes derived from
				\{{N}$_{\textrm{3}}${O}\}-donor {Schiff} base ligands},}\ }\href {\doibase
		10.1039/C7NJ00158D} {\bibfield  {journal} {\bibinfo  {journal} {New J.
				Chem.}\ }\textbf {\bibinfo {volume} {41}},\ \bibinfo {pages} {3143} (\bibinfo
		{year} {2017})}\BibitemShut {NoStop}%
	\bibitem [{\citenamefont {Das}\ \emph {et~al.}(2017)\citenamefont {Das},
		\citenamefont {Bhattacharya}, \citenamefont {Giri},\ and\ \citenamefont
		{Ghosh}}]{das_2017}%
	\BibitemOpen
	\bibfield  {author} {\bibinfo {author} {\bibfnamefont {Avijit}\ \bibnamefont
			{Das}}, \bibinfo {author} {\bibfnamefont {Kisholoy}\ \bibnamefont
			{Bhattacharya}}, \bibinfo {author} {\bibfnamefont {Sanjib}\ \bibnamefont
			{Giri}}, \ and\ \bibinfo {author} {\bibfnamefont {Ashutosh}\ \bibnamefont
			{Ghosh}},\ }\bibfield  {title} {\enquote {\bibinfo {title} {Synthesis,
				crystal structure and magnetic properties of a dinuclear and a trinuclear
				{Ni}({II}) complexes derived from tetradentate {ONNO} donor {Mannich} base
				ligands},}\ }\href {\doibase 10.1016/j.poly.2017.06.022} {\bibfield
		{journal} {\bibinfo  {journal} {Polyhedron}\ }\textbf {\bibinfo {volume}
			{134}},\ \bibinfo {pages} {295} (\bibinfo {year} {2017})}\BibitemShut
	{NoStop}%
	\bibitem [{\citenamefont {Woods}\ \emph {et~al.}(2017)\citenamefont {Woods},
		\citenamefont {Stout}, \citenamefont {Dolinar}, \citenamefont {Vignesh},
		\citenamefont {Ballesteros-Rivas}, \citenamefont {Achim},\ and\ \citenamefont
		{Dunbar}}]{woods_2017}%
	\BibitemOpen
	\bibfield  {author} {\bibinfo {author} {\bibfnamefont {Toby~J.}\ \bibnamefont
			{Woods}}, \bibinfo {author} {\bibfnamefont {Heather~D.}\ \bibnamefont
			{Stout}}, \bibinfo {author} {\bibfnamefont {Brian~S.}\ \bibnamefont
			{Dolinar}}, \bibinfo {author} {\bibfnamefont {Kuduva~R.}\ \bibnamefont
			{Vignesh}}, \bibinfo {author} {\bibfnamefont {Maria~F.}\ \bibnamefont
			{Ballesteros-Rivas}}, \bibinfo {author} {\bibfnamefont {Catalina}\
			\bibnamefont {Achim}}, \ and\ \bibinfo {author} {\bibfnamefont {Kim~R.}\
			\bibnamefont {Dunbar}},\ }\bibfield  {title} {\enquote {\bibinfo {title}
			{Strong {Ferromagnetic} {Exchange} {Coupling} {Mediated} by a {Bridging}
				{Tetrazine} {Radical} in a {Dinuclear} {Nickel} {Complex}},}\ }\href
	{\doibase 10.1021/acs.inorgchem.7b01812} {\bibfield  {journal} {\bibinfo
			{journal} {Inorg. Chem.}\ }\textbf {\bibinfo {volume} {56}},\ \bibinfo
		{pages} {12094} (\bibinfo {year} {2017})}\BibitemShut {NoStop}%
	\bibitem [{\citenamefont {Goodenough}(1955)}]{goodenough_1955}%
	\BibitemOpen
	\bibfield  {author} {\bibinfo {author} {\bibfnamefont {John~B.}\ \bibnamefont
			{Goodenough}},\ }\bibfield  {title} {\enquote {\bibinfo {title} {Theory of
				the {Role} of {Covalence} in the {Perovskite}-{Type} {Manganites}
				[{La},{M}({II})]{Mn}{O}$_3$},}\ }\href {\doibase 10.1103/PhysRev.100.564}
	{\bibfield  {journal} {\bibinfo  {journal} {Physical Review}\ }\textbf
		{\bibinfo {volume} {100}},\ \bibinfo {pages} {564--573} (\bibinfo {year}
		{1955})}\BibitemShut {NoStop}%
	\bibitem [{\citenamefont {DeFotis}\ \emph {et~al.}(1990)\citenamefont
		{DeFotis}, \citenamefont {Remy},\ and\ \citenamefont
		{Scherrer}}]{defotis_1990}%
	\BibitemOpen
	\bibfield  {author} {\bibinfo {author} {\bibfnamefont {G.~C.}\ \bibnamefont
			{DeFotis}}, \bibinfo {author} {\bibfnamefont {E.~D.}\ \bibnamefont {Remy}}, \
		and\ \bibinfo {author} {\bibfnamefont {C.~W.}\ \bibnamefont {Scherrer}},\
	}\bibfield  {title} {\enquote {\bibinfo {title} {Magnetic and structural
				properties of {Mn}({SCN})$_2$({CH}$_3${OH})$_2$: {A} quasi-two-dimensional
				{Heisenberg} antiferromagnet},}\ }\href {\doibase 10.1103/PhysRevB.41.9074}
	{\bibfield  {journal} {\bibinfo  {journal} {Phys. Rev. B}\ }\textbf {\bibinfo
			{volume} {41}},\ \bibinfo {pages} {9074} (\bibinfo {year}
		{1990})}\BibitemShut {NoStop}%
	\bibitem [{\citenamefont {Law}\ \emph {et~al.}(2000)\citenamefont {Law},
		\citenamefont {Kampf},\ and\ \citenamefont {Pecoraro}}]{law_2000}%
	\BibitemOpen
	\bibfield  {author} {\bibinfo {author} {\bibfnamefont {Neil~A.}\ \bibnamefont
			{Law}}, \bibinfo {author} {\bibfnamefont {Jeff~W.}\ \bibnamefont {Kampf}}, \
		and\ \bibinfo {author} {\bibfnamefont {Vincent~L.}\ \bibnamefont
			{Pecoraro}},\ }\bibfield  {title} {\enquote {\bibinfo {title} {{A
					magneto-structural correlation between the {Heisenberg} constant, ${J}$, and
					the {Mn}-{O}-{Mn} angle in [{Mn$^{\mathrm{IV}}$}($\mu$-{O})]$_2$ dimers}},}\
	}\href {\doibase 10.1016/S0020-1693(99)00430-2} {\bibfield  {journal}
		{\bibinfo  {journal} {Inorganica Chim. Acta}\ }\textbf {\bibinfo {volume}
			{297}},\ \bibinfo {pages} {252} (\bibinfo {year} {2000})}\BibitemShut
	{NoStop}%
	\bibitem [{\citenamefont {Han}\ \emph {et~al.}(2004)\citenamefont {Han},
		\citenamefont {Ozaki},\ and\ \citenamefont {Yu}}]{han_2004}%
	\BibitemOpen
	\bibfield  {author} {\bibinfo {author} {\bibfnamefont {Myung~Joon}\
			\bibnamefont {Han}}, \bibinfo {author} {\bibfnamefont {Taisuke}\ \bibnamefont
			{Ozaki}}, \ and\ \bibinfo {author} {\bibfnamefont {Jaejun}\ \bibnamefont
			{Yu}},\ }\bibfield  {title} {\enquote {\bibinfo {title} {Electronic
				structure, magnetic interactions, and the role of ligands in
				{Mn}$_n$($n$=4,12) single-molecule magnets},}\ }\href {\doibase
		10.1103/PhysRevB.70.184421} {\bibfield  {journal} {\bibinfo  {journal}
			{Physical Review B}\ }\textbf {\bibinfo {volume} {70}},\ \bibinfo {pages}
		{184421} (\bibinfo {year} {2004})}\BibitemShut {NoStop}%
	\bibitem [{\citenamefont {Perks}\ \emph {et~al.}(2012)\citenamefont {Perks},
		\citenamefont {Johnson}, \citenamefont {Martin}, \citenamefont {Chapon},\
		and\ \citenamefont {Radaelli}}]{perks_2012}%
	\BibitemOpen
	\bibfield  {author} {\bibinfo {author} {\bibfnamefont {N.J.}\ \bibnamefont
			{Perks}}, \bibinfo {author} {\bibfnamefont {R.D.}\ \bibnamefont {Johnson}},
		\bibinfo {author} {\bibfnamefont {C.}~\bibnamefont {Martin}}, \bibinfo
		{author} {\bibfnamefont {L.C.}\ \bibnamefont {Chapon}}, \ and\ \bibinfo
		{author} {\bibfnamefont {P.G.}\ \bibnamefont {Radaelli}},\ }\bibfield
	{title} {\enquote {\bibinfo {title} {{Magneto-orbital helices as a route to
					coupling magnetism and ferroelectricity in multiferroic CaMn$_7$O$_{12}$}},}\
	}\href {http://www.nature.com/articles/ncomms2294} {\bibfield  {journal}
		{\bibinfo  {journal} {Nature Communications}\ }\textbf {\bibinfo {volume}
			{3}} (\bibinfo {year} {2012})}\BibitemShut {NoStop}%
	\bibitem [{\citenamefont {Gupta}\ and\ \citenamefont
		{Rajaraman}(2016)}]{gupta_2016}%
	\BibitemOpen
	\bibfield  {author} {\bibinfo {author} {\bibfnamefont {Tulika}\ \bibnamefont
			{Gupta}}\ and\ \bibinfo {author} {\bibfnamefont {Gopalan}\ \bibnamefont
			{Rajaraman}},\ }\bibfield  {title} {\enquote {\bibinfo {title} {Modelling
				spin {Hamiltonian} parameters of molecular nanomagnets},}\ }\href {\doibase
		10.1039/C6CC01251E} {\bibfield  {journal} {\bibinfo  {journal} {Chemical
				Communications}\ }\textbf {\bibinfo {volume} {52}},\ \bibinfo {pages}
		{8972--9008} (\bibinfo {year} {2016})}\BibitemShut {NoStop}%
	\bibitem [{\citenamefont {H\"{a}nninen}\ \emph {et~al.}(2018)\citenamefont
		{H\"{a}nninen}, \citenamefont {Mota}, \citenamefont {Sillanp\"{a}\"{a}},
		\citenamefont {Dey}, \citenamefont {Velmurugan}, \citenamefont {Rajaraman},\
		and\ \citenamefont {Colacio}}]{hanninen_2018}%
	\BibitemOpen
	\bibfield  {author} {\bibinfo {author} {\bibfnamefont {Mikko~M.}\
			\bibnamefont {H\"{a}nninen}}, \bibinfo {author} {\bibfnamefont {Antonio~J.}\
			\bibnamefont {Mota}}, \bibinfo {author} {\bibfnamefont {Reijo}\ \bibnamefont
			{Sillanp\"{a}\"{a}}}, \bibinfo {author} {\bibfnamefont {Sourav}\ \bibnamefont
			{Dey}}, \bibinfo {author} {\bibfnamefont {Gunasekaran}\ \bibnamefont
			{Velmurugan}}, \bibinfo {author} {\bibfnamefont {Gopalan}\ \bibnamefont
			{Rajaraman}}, \ and\ \bibinfo {author} {\bibfnamefont {Enrique}\ \bibnamefont
			{Colacio}},\ }\bibfield  {title} {\enquote {\bibinfo {title}
			{Magneto-{Structural} {Properties} and {Theoretical} {Studies} of a {Family}
				of {Simple} {Heterodinuclear} {Phenoxide}/{Alkoxide} {Bridged}
				{Mn}$^{\textrm{{iii}}}${Ln}$^{\textrm{{iii}}}$ {Complexes}: {On} the {Nature}
				of the {Magnetic} {Exchange} and {Magnetic} {Anisotropy}},}\ }\href {\doibase
		10.1021/acs.inorgchem.7b02917} {\bibfield  {journal} {\bibinfo  {journal}
			{Inorg. Chem.}\ }\textbf {\bibinfo {volume} {57}},\ \bibinfo {pages} {3683}
		(\bibinfo {year} {2018})}\BibitemShut {NoStop}%
	\bibitem [{\citenamefont {Georgiev}\ and\ \citenamefont
		{Chamati}(2018)}]{georgiev_2018}%
	\BibitemOpen
	\bibfield  {author} {\bibinfo {author} {\bibfnamefont {M.}~\bibnamefont
			{Georgiev}}\ and\ \bibinfo {author} {\bibfnamefont {H.}~\bibnamefont
			{Chamati}},\ }\bibfield  {title} {\enquote {\bibinfo {title} {A systematic
				approach to determine the spectral characteristics of molecular magnets},}\
	}\href {http://arxiv.org/abs/1805.01382} {\bibfield  {journal} {\bibinfo
			{journal} {arXiv:1805.01382 [cond-mat]}\ } (\bibinfo {year}
		{2018})}\BibitemShut {NoStop}%
	\bibitem [{\citenamefont {Fleming}(2009)}]{fleming_molecular_2009}%
	\BibitemOpen
	\bibfield  {author} {\bibinfo {author} {\bibfnamefont {Ian}\ \bibnamefont
			{Fleming}},\ }\href {\doibase 10.1002/9780470684306} {\emph {\bibinfo {title}
			{Molecular {Orbitals} and {Organic} {Chemical} {Reactions}}}}\ (\bibinfo
	{publisher} {Wiley},\ \bibinfo {address} {Chichester, UK},\ \bibinfo {year}
	{2009})\BibitemShut {NoStop}%
	\bibitem [{\citenamefont {Lennard-Jones}(1929)}]{lennard_1929}%
	\BibitemOpen
	\bibfield  {author} {\bibinfo {author} {\bibfnamefont {J.~E.}\ \bibnamefont
			{Lennard-Jones}},\ }\bibfield  {title} {\enquote {\bibinfo {title} {The
				electronic structure of some diatomic molecules},}\ }\href {\doibase
		10.1039/tf9292500668} {\bibfield  {journal} {\bibinfo  {journal} {Trans.
				Faraday Soc.,}\ }\textbf {\bibinfo {volume} {25}},\ \bibinfo {pages} {668}
		(\bibinfo {year} {1929})}\BibitemShut {NoStop}%
	\bibitem [{\citenamefont {Mulliken}(1939)}]{mulliken_1939}%
	\BibitemOpen
	\bibfield  {author} {\bibinfo {author} {\bibfnamefont {Robert~S.}\
			\bibnamefont {Mulliken}},\ }\bibfield  {title} {\enquote {\bibinfo {title}
			{Intensities of {Electronic} {Transitions} in {Molecular} {Spectra} {II}.
				{Charge}-{Transfer} {Spectra}},}\ }\href {\doibase 10.1063/1.1750319}
	{\bibfield  {journal} {\bibinfo  {journal} {J. Chem. Phys.}\ }\textbf
		{\bibinfo {volume} {7}},\ \bibinfo {pages} {20} (\bibinfo {year}
		{1939})}\BibitemShut {NoStop}%
	\bibitem [{\citenamefont {Lennard-Jones}(1949)}]{lennard-jones_1949}%
	\BibitemOpen
	\bibfield  {author} {\bibinfo {author} {\bibfnamefont {J.}~\bibnamefont
			{Lennard-Jones}},\ }\bibfield  {title} {\enquote {\bibinfo {title} {The
				{Molecular} {Orbital} {Theory} of {Chemical} {Valency}. {I}. {The}
				{Determination} of {Molecular} {Orbitals}},}\ }\href {\doibase
		10.1098/rspa.1949.0083} {\bibfield  {journal} {\bibinfo  {journal} {Proc. R.
				Soc. London, Ser. A}\ }\textbf {\bibinfo {volume} {198}},\ \bibinfo {pages}
		{1} (\bibinfo {year} {1949})}\BibitemShut {NoStop}%
	\bibitem [{\citenamefont {Pople}\ \emph {et~al.}(1965)\citenamefont {Pople},
		\citenamefont {Santry},\ and\ \citenamefont {Segal}}]{pople_1965}%
	\BibitemOpen
	\bibfield  {author} {\bibinfo {author} {\bibfnamefont {J.~A.}\ \bibnamefont
			{Pople}}, \bibinfo {author} {\bibfnamefont {D.~P.}\ \bibnamefont {Santry}}, \
		and\ \bibinfo {author} {\bibfnamefont {G.~A.}\ \bibnamefont {Segal}},\
	}\bibfield  {title} {\enquote {\bibinfo {title} {Approximate
				{Self}-{Consistent} {Molecular} {Orbital} {Theory}. {I}. {Invariant}
				{Procedures}},}\ }\href {\doibase 10.1063/1.1701475} {\bibfield  {journal}
		{\bibinfo  {journal} {J. Chem. Phys.}\ }\textbf {\bibinfo {volume} {43}},\
		\bibinfo {pages} {S129} (\bibinfo {year} {1965})}\BibitemShut {NoStop}%
	\bibitem [{\citenamefont {Pople}\ and\ \citenamefont
		{Segal}(1966)}]{pople_1966}%
	\BibitemOpen
	\bibfield  {author} {\bibinfo {author} {\bibfnamefont {J.~A.}\ \bibnamefont
			{Pople}}\ and\ \bibinfo {author} {\bibfnamefont {G.~A.}\ \bibnamefont
			{Segal}},\ }\bibfield  {title} {\enquote {\bibinfo {title} {Approximate
				{Self}-{Consistent} {Molecular} {Orbital} {Theory}. {III}. {CNDO} {Results}
				for {AB}$_{\textrm{2}}$ and {AB}$_{\textrm{3}}$ {Systems}},}\ }\href
	{\doibase 10.1063/1.1727227} {\bibfield  {journal} {\bibinfo  {journal} {J.
				Chem. Phys.}\ }\textbf {\bibinfo {volume} {44}},\ \bibinfo {pages} {3289}
		(\bibinfo {year} {1966})}\BibitemShut {NoStop}%
	\bibitem [{\citenamefont {Pople}\ \emph {et~al.}(1967)\citenamefont {Pople},
		\citenamefont {Beveridge},\ and\ \citenamefont {Dobosh}}]{pople_1967}%
	\BibitemOpen
	\bibfield  {author} {\bibinfo {author} {\bibfnamefont {J.~A.}\ \bibnamefont
			{Pople}}, \bibinfo {author} {\bibfnamefont {D.~L.}\ \bibnamefont
			{Beveridge}}, \ and\ \bibinfo {author} {\bibfnamefont {P.~A.}\ \bibnamefont
			{Dobosh}},\ }\bibfield  {title} {\enquote {\bibinfo {title} {Approximate
				{Self}-{Consistent} {Molecular}-{Orbital} {Theory}. {V}. {Intermediate}
				{Neglect} of {Differential} {Overlap}},}\ }\href {\doibase 10.1063/1.1712233}
	{\bibfield  {journal} {\bibinfo  {journal} {J. Chem. Phys.}\ }\textbf
		{\bibinfo {volume} {47}},\ \bibinfo {pages} {2026} (\bibinfo {year}
		{1967})}\BibitemShut {NoStop}%
	\bibitem [{\citenamefont {England}\ \emph {et~al.}(1971)\citenamefont
		{England}, \citenamefont {Salmon},\ and\ \citenamefont
		{Ruedenberg}}]{england_1971}%
	\BibitemOpen
	\bibfield  {author} {\bibinfo {author} {\bibfnamefont {Walter}\ \bibnamefont
			{England}}, \bibinfo {author} {\bibfnamefont {Lydia~S.}\ \bibnamefont
			{Salmon}}, \ and\ \bibinfo {author} {\bibfnamefont {Klaus}\ \bibnamefont
			{Ruedenberg}},\ }\bibfield  {title} {\enquote {\bibinfo {title} {Localized
				molecular orbitals: {A} bridge between chemical intuition and molecular
				quantum mechanics},}\ }in\ \href {\doibase 10.1007/BFb0051440} {\emph
		{\bibinfo {booktitle} {Molecular {Orbitals}}}},\ Vol.\ \bibinfo {volume}
	{23/1}\ (\bibinfo  {publisher} {Springer-Verlag},\ \bibinfo {address}
	{Berlin/Heidelberg},\ \bibinfo {year} {1971})\ p.~\bibinfo {pages}
	{31}\BibitemShut {NoStop}%
	\bibitem [{\citenamefont {Reed}\ and\ \citenamefont
		{Weinhold}(1985)}]{reed_1985}%
	\BibitemOpen
	\bibfield  {author} {\bibinfo {author} {\bibfnamefont {Alan~E.}\ \bibnamefont
			{Reed}}\ and\ \bibinfo {author} {\bibfnamefont {Frank}\ \bibnamefont
			{Weinhold}},\ }\bibfield  {title} {\enquote {\bibinfo {title} {Natural
				localized molecular orbitals},}\ }\href {\doibase 10.1063/1.449360}
	{\bibfield  {journal} {\bibinfo  {journal} {J. Chem. Phys.}\ }\textbf
		{\bibinfo {volume} {83}},\ \bibinfo {pages} {1736} (\bibinfo {year}
		{1985})}\BibitemShut {NoStop}%
	\bibitem [{\citenamefont {Slater}(1929)}]{slater_1929}%
	\BibitemOpen
	\bibfield  {author} {\bibinfo {author} {\bibfnamefont {J.~C.}\ \bibnamefont
			{Slater}},\ }\bibfield  {title} {\enquote {\bibinfo {title} {The {Theory} of
				{Complex} {Spectra}},}\ }\href {\doibase 10.1103/PhysRev.34.1293} {\bibfield
		{journal} {\bibinfo  {journal} {Phys. Rev.}\ }\textbf {\bibinfo {volume}
			{34}},\ \bibinfo {pages} {1293} (\bibinfo {year} {1929})}\BibitemShut
	{NoStop}%
	\bibitem [{\citenamefont {Slater}(1930)}]{slater_1930}%
	\BibitemOpen
	\bibfield  {author} {\bibinfo {author} {\bibfnamefont {J.~C.}\ \bibnamefont
			{Slater}},\ }\bibfield  {title} {\enquote {\bibinfo {title} {Note on
				{Hartree}'s {Method}},}\ }\href {\doibase 10.1103/PhysRev.35.210.2}
	{\bibfield  {journal} {\bibinfo  {journal} {Phys. Rev.}\ }\textbf {\bibinfo
			{volume} {35}},\ \bibinfo {pages} {210} (\bibinfo {year} {1930})}\BibitemShut
	{NoStop}%
	\bibitem [{\citenamefont {Hartree}(1928)}]{hartree_1928}%
	\BibitemOpen
	\bibfield  {author} {\bibinfo {author} {\bibfnamefont {D.~R.}\ \bibnamefont
			{Hartree}},\ }\bibfield  {title} {\enquote {\bibinfo {title} {The {Wave}
				{Mechanics} of an {Atom} with a non-{Coulomb} {Central} {Field}. {Part}
				{III}. {Term} {Values} and {Intensities} in {Series} in {Optical}
				{Spectra}},}\ }\href {\doibase 10.1017/S0305004100015954} {\bibfield
		{journal} {\bibinfo  {journal} {Math. Proc. Camb. Philos. Soc.}\ }\textbf
		{\bibinfo {volume} {24}},\ \bibinfo {pages} {426} (\bibinfo {year}
		{1928})}\BibitemShut {NoStop}%
	\bibitem [{\citenamefont {Tsuneda}(2014)}]{tsuneda_2014}%
	\BibitemOpen
	\bibfield  {author} {\bibinfo {author} {\bibfnamefont {Takao}\ \bibnamefont
			{Tsuneda}},\ }\bibfield  {title} {\enquote {\bibinfo {title} {Hartree-{Fock}
				{Method}},}\ }in\ \href {\doibase 10.1007/978-4-431-54825-6_2} {\emph
		{\bibinfo {booktitle} {Density {Functional} {Theory} in {Quantum}
				{Chemistry}}}}\ (\bibinfo  {publisher} {Springer Japan},\ \bibinfo {address}
	{Tokyo},\ \bibinfo {year} {2014})\ p.~\bibinfo {pages} {35}\BibitemShut
	{NoStop}%
	\bibitem [{\citenamefont {Anderson}(1950)}]{anderson_1950}%
	\BibitemOpen
	\bibfield  {author} {\bibinfo {author} {\bibfnamefont {P.~W.}\ \bibnamefont
			{Anderson}},\ }\bibfield  {title} {\enquote {\bibinfo {title}
			{Antiferromagnetism. {Theory} of {Superexchange} {Interaction}},}\ }\href
	{\doibase 10.1103/PhysRev.79.350} {\bibfield  {journal} {\bibinfo  {journal}
			{Phys. Rev.}\ }\textbf {\bibinfo {volume} {79}},\ \bibinfo {pages} {350}
		(\bibinfo {year} {1950})}\BibitemShut {NoStop}%
	\bibitem [{\citenamefont {Kanamori}(1959)}]{kanamori_1959}%
	\BibitemOpen
	\bibfield  {author} {\bibinfo {author} {\bibfnamefont {Junjiro}\ \bibnamefont
			{Kanamori}},\ }\bibfield  {title} {\enquote {\bibinfo {title} {Superexchange
				interaction and symmetry properties of electron orbitals},}\ }\href {\doibase
		10.1016/0022-3697(59)90061-7} {\bibfield  {journal} {\bibinfo  {journal} {J.
				Phys. Chem. Solids}\ }\textbf {\bibinfo {volume} {10}},\ \bibinfo {pages}
		{87} (\bibinfo {year} {1959})}\BibitemShut {NoStop}%
	\bibitem [{\citenamefont {Anderson}(1959)}]{anderson_1959}%
	\BibitemOpen
	\bibfield  {author} {\bibinfo {author} {\bibfnamefont {P.~W.}\ \bibnamefont
			{Anderson}},\ }\bibfield  {title} {\enquote {\bibinfo {title} {New {Approach}
				to the {Theory} of {Superexchange} {Interactions}},}\ }\href {\doibase
		10.1103/PhysRev.115.2} {\bibfield  {journal} {\bibinfo  {journal} {Phys.
				Rev.}\ }\textbf {\bibinfo {volume} {115}},\ \bibinfo {pages} {2} (\bibinfo
		{year} {1959})}\BibitemShut {NoStop}%
	\bibitem [{\citenamefont {Heisenberg}(1926)}]{heisenberg_1926}%
	\BibitemOpen
	\bibfield  {author} {\bibinfo {author} {\bibfnamefont {W.}~\bibnamefont
			{Heisenberg}},\ }\bibfield  {title} {\enquote {\bibinfo {title}
			{Mehrk\"{o}rperproblem und {Resonanz} in der {Quantenmechanik}},}\ }\href
	{\doibase 10.1007/BF01397160} {\bibfield  {journal} {\bibinfo  {journal}
			{Zeitschrift f\"{u}r Physik}\ }\textbf {\bibinfo {volume} {38}},\ \bibinfo
		{pages} {411} (\bibinfo {year} {1926})}\BibitemShut {NoStop}%
	\bibitem [{\citenamefont {Hubbard}(1963)}]{hubbard_1963}%
	\BibitemOpen
	\bibfield  {author} {\bibinfo {author} {\bibfnamefont {J.}~\bibnamefont
			{Hubbard}},\ }\bibfield  {title} {\enquote {\bibinfo {title} {Electron
				correlations in narrow energy bands},}\ }\href
	{http://www.jstor.org/stable/2414761} {\bibfield  {journal} {\bibinfo
			{journal} {Proc. R. Soc. A}\ }\textbf {\bibinfo {volume} {276}},\ \bibinfo
		{pages} {238--257} (\bibinfo {year} {1963})}\BibitemShut {NoStop}%
	\bibitem [{\citenamefont {Hubbard}(1964)}]{hubbard_1964}%
	\BibitemOpen
	\bibfield  {author} {\bibinfo {author} {\bibfnamefont {J.}~\bibnamefont
			{Hubbard}},\ }\bibfield  {title} {\enquote {\bibinfo {title} {Electron
				{Correlations} in {Narrow} {Energy} {Bands}. {III}. {An} {Improved}
				{Solution}},}\ }\href {\doibase 10.1098/rspa.1964.0190} {\bibfield  {journal}
		{\bibinfo  {journal} {Proc. R. Soc. A}\ }\textbf {\bibinfo {volume} {281}},\
		\bibinfo {pages} {401--419} (\bibinfo {year} {1964})}\BibitemShut {NoStop}%
	\bibitem [{\citenamefont {M\"{u}ller}\ \emph {et~al.}(2000)\citenamefont
		{M\"{u}ller}, \citenamefont {Beugholt}, \citenamefont {K\"{o}gerler},
		\citenamefont {B\"{o}gge}, \citenamefont {Bud'ko},\ and\ \citenamefont
		{Luban}}]{muller_2000}%
	\BibitemOpen
	\bibfield  {author} {\bibinfo {author} {\bibfnamefont {Achim}\ \bibnamefont
			{M\"{u}ller}}, \bibinfo {author} {\bibfnamefont {Christian}\ \bibnamefont
			{Beugholt}}, \bibinfo {author} {\bibfnamefont {Paul}\ \bibnamefont
			{K\"{o}gerler}}, \bibinfo {author} {\bibfnamefont {Hartmut}\ \bibnamefont
			{B\"{o}gge}}, \bibinfo {author} {\bibfnamefont {Sergey}\ \bibnamefont
			{Bud'ko}}, \ and\ \bibinfo {author} {\bibfnamefont {Marshall}\ \bibnamefont
			{Luban}},\ }\bibfield  {title} {\enquote {\bibinfo {title}
			{{[Mo$^{\textrm{V}}_{12}$O$_{30}$($\mu_2$-OH)$_{10}$H$_2$\{Ni$^{\textrm{II}}$(H$_2$O)$_3$\}$_4$],
					a Highly Symmetrical $\varepsilon$-Keggin Unit Capped with Four
					Ni$^{\textrm{II}}$ Centers: Synthesis and Magnetism}},}\ }\href {\doibase
		10.1021/ic0005285} {\bibfield  {journal} {\bibinfo  {journal} {Inorg. Chem.}\
		}\textbf {\bibinfo {volume} {39}},\ \bibinfo {pages} {5176} (\bibinfo {year}
		{2000})}\BibitemShut {NoStop}%
	\bibitem [{\citenamefont {Marzari}\ \emph {et~al.}(2012)\citenamefont
		{Marzari}, \citenamefont {Mostofi}, \citenamefont {Yates}, \citenamefont
		{Souza},\ and\ \citenamefont {Vanderbilt}}]{marzari_2012}%
	\BibitemOpen
	\bibfield  {author} {\bibinfo {author} {\bibfnamefont {Nicola}\ \bibnamefont
			{Marzari}}, \bibinfo {author} {\bibfnamefont {Arash~A.}\ \bibnamefont
			{Mostofi}}, \bibinfo {author} {\bibfnamefont {Jonathan~R.}\ \bibnamefont
			{Yates}}, \bibinfo {author} {\bibfnamefont {Ivo}\ \bibnamefont {Souza}}, \
		and\ \bibinfo {author} {\bibfnamefont {David}\ \bibnamefont {Vanderbilt}},\
	}\bibfield  {title} {\enquote {\bibinfo {title} {{Maximally localized Wannier
					functions: Theory and applications}},}\ }\href {\doibase
		10.1103/RevModPhys.84.1419} {\bibfield  {journal} {\bibinfo  {journal} {Rev.
				Mod. Phys.}\ }\textbf {\bibinfo {volume} {84}},\ \bibinfo {pages} {1419}
		(\bibinfo {year} {2012})}\BibitemShut {NoStop}%
	\bibitem [{\citenamefont {Anderson}(1963)}]{anderson_1963}%
	\BibitemOpen
	\bibfield  {author} {\bibinfo {author} {\bibfnamefont {Philip~W.}\
			\bibnamefont {Anderson}},\ }\bibfield  {title} {\enquote {\bibinfo {title}
			{Theory of {Magnetic} {Exchange} {Interactions}:{Exchange} in {Insulators}
				and {Semiconductors}},}\ }in\ \href {\doibase 10.1016/S0081-1947(08)60260-X}
	{\emph {\bibinfo {booktitle} {Solid {State} {Physics}}}},\ Vol.~\bibinfo
	{volume} {14}\ (\bibinfo  {publisher} {Elsevier},\ \bibinfo {year} {1963})\
	p.~\bibinfo {pages} {99}\BibitemShut {NoStop}%
	\bibitem [{\citenamefont {Matsuda}\ \emph {et~al.}(2005)\citenamefont
		{Matsuda}, \citenamefont {Kakurai}, \citenamefont {Belik}, \citenamefont
		{Azuma}, \citenamefont {Takano},\ and\ \citenamefont
		{Fujita}}]{matsuda_magnetic_2005}%
	\BibitemOpen
	\bibfield  {author} {\bibinfo {author} {\bibfnamefont {M.}~\bibnamefont
			{Matsuda}}, \bibinfo {author} {\bibfnamefont {K.}~\bibnamefont {Kakurai}},
		\bibinfo {author} {\bibfnamefont {A.~A.}\ \bibnamefont {Belik}}, \bibinfo
		{author} {\bibfnamefont {M.}~\bibnamefont {Azuma}}, \bibinfo {author}
		{\bibfnamefont {M.}~\bibnamefont {Takano}}, \ and\ \bibinfo {author}
		{\bibfnamefont {M.}~\bibnamefont {Fujita}},\ }\bibfield  {title} {\enquote
		{\bibinfo {title} {{Magnetic excitations from the linear Heisenberg
					antiferromagnetic spin trimer system A$_3$Cu$_3$(PO$_4)_4$ (A = Ca, Sr, and
					Pb)}},}\ }\href {\doibase 10.1103/PhysRevB.71.144411} {\bibfield  {journal}
		{\bibinfo  {journal} {Phys. Rev. B}\ }\textbf {\bibinfo {volume} {71}},\
		\bibinfo {pages} {144411} (\bibinfo {year} {2005})}\BibitemShut {NoStop}%
	\bibitem [{\citenamefont {Podlesnyak}\ \emph {et~al.}(2007)\citenamefont
		{Podlesnyak}, \citenamefont {Pomjakushin}, \citenamefont {Pomjakushina},
		\citenamefont {Conder},\ and\ \citenamefont
		{Furrer}}]{podlesnyak_magnetic_2007}%
	\BibitemOpen
	\bibfield  {author} {\bibinfo {author} {\bibfnamefont {A.}~\bibnamefont
			{Podlesnyak}}, \bibinfo {author} {\bibfnamefont {V.}~\bibnamefont
			{Pomjakushin}}, \bibinfo {author} {\bibfnamefont {E.}~\bibnamefont
			{Pomjakushina}}, \bibinfo {author} {\bibfnamefont {K.}~\bibnamefont
			{Conder}}, \ and\ \bibinfo {author} {\bibfnamefont {A.}~\bibnamefont
			{Furrer}},\ }\bibfield  {title} {\enquote {\bibinfo {title} {{Magnetic
					excitations in the spin-trimer compounds Ca$_3$Cu$_{3-x}$Ni$_x$(PO$_4)_4$
					($x=0 , 1 , 2$)}},}\ }\href {\doibase 10.1103/PhysRevB.76.064420} {\bibfield
		{journal} {\bibinfo  {journal} {Phys. Rev. B}\ }\textbf {\bibinfo {volume}
			{76}},\ \bibinfo {pages} {064420} (\bibinfo {year} {2007})}\BibitemShut
	{NoStop}%
	\bibitem [{\citenamefont {Nehrkorn}\ \emph {et~al.}(2010)\citenamefont
		{Nehrkorn}, \citenamefont {H\"{o}ck}, \citenamefont {Br\"{u}ger},
		\citenamefont {Mutka}, \citenamefont {Schnack},\ and\ \citenamefont
		{Waldmann}}]{nehrkorn_inelastic_2010}%
	\BibitemOpen
	\bibfield  {author} {\bibinfo {author} {\bibfnamefont {J.}~\bibnamefont
			{Nehrkorn}}, \bibinfo {author} {\bibfnamefont {M.}~\bibnamefont {H\"{o}ck}},
		\bibinfo {author} {\bibfnamefont {M.}~\bibnamefont {Br\"{u}ger}}, \bibinfo
		{author} {\bibfnamefont {H.}~\bibnamefont {Mutka}}, \bibinfo {author}
		{\bibfnamefont {J.}~\bibnamefont {Schnack}}, \ and\ \bibinfo {author}
		{\bibfnamefont {O.}~\bibnamefont {Waldmann}},\ }\bibfield  {title} {\enquote
		{\bibinfo {title} {{Inelastic neutron scattering study and Hubbard model
					description of the antiferromagnetic tetrahedral molecule
					Ni$_4$Mo$_{12}$}},}\ }\href {\doibase 10.1140/epjb/e2010-00028-3} {\bibfield
		{journal} {\bibinfo  {journal} {Eur. Phys. J. B}\ }\textbf {\bibinfo {volume}
			{73}},\ \bibinfo {pages} {515} (\bibinfo {year} {2010})}\BibitemShut
	{NoStop}%
	\bibitem [{\citenamefont {Furrer}\ \emph {et~al.}(2010)\citenamefont {Furrer},
		\citenamefont {Kr\"{a}mer}, \citenamefont {Str\"{a}ssle}, \citenamefont
		{Biner}, \citenamefont {Hauser},\ and\ \citenamefont
		{G\"{u}del}}]{furrer_magnetic_PRB_2010}%
	\BibitemOpen
	\bibfield  {author} {\bibinfo {author} {\bibfnamefont {A.}~\bibnamefont
			{Furrer}}, \bibinfo {author} {\bibfnamefont {K.~W.}\ \bibnamefont
			{Kr\"{a}mer}}, \bibinfo {author} {\bibfnamefont {Th.}\ \bibnamefont
			{Str\"{a}ssle}}, \bibinfo {author} {\bibfnamefont {D.}~\bibnamefont {Biner}},
		\bibinfo {author} {\bibfnamefont {J.}~\bibnamefont {Hauser}}, \ and\ \bibinfo
		{author} {\bibfnamefont {H.~U.}\ \bibnamefont {G\"{u}del}},\ }\bibfield
	{title} {\enquote {\bibinfo {title} {{Magnetic and neutron spectroscopic
					properties of the tetrameric nickel compound
					[Mo$_{12}${O}$_{28}$($\mu_2$-{OH})$_9$($\mu_2$-OH)$_3$\{Ni(H$_2$O)$_3$\}$_4$]
					$\cdot$ 13H$_2$O}},}\ }\href {\doibase 10.1103/PhysRevB.81.214437} {\bibfield
		{journal} {\bibinfo  {journal} {Phys. Rev. B}\ }\textbf {\bibinfo {volume}
			{81}},\ \bibinfo {pages} {214437} (\bibinfo {year} {2010})}\BibitemShut
	{NoStop}%
\end{thebibliography}
\end{document}